\documentclass[aps,pre,10pt,twocolumn,longbibliography]{revtex4-1}
\usepackage{microtype}
\usepackage{graphicx}
\usepackage[colorlinks,linkcolor=blue,urlcolor=blue,citecolor=blue]{hyperref}
\usepackage{subcaption}
\usepackage{times}
\usepackage{epsfig,amsopn}
\usepackage{graphicx}
\usepackage{bm,amsmath,amssymb}
\usepackage{natbib}
\usepackage{braket}
\usepackage{float}
\usepackage{wasysym}
\usepackage{enumerate}
\usepackage{hyperref}
\usepackage{comment}
\usepackage{tikz}
\allowdisplaybreaks[1]
\usepackage[dvipsnames]{xcolor}

\def\be{\begin{equation}}
\def\ee{\end{equation}}
\def\bse{\begin{subequations}}
\def\ese{\end{subequations}}
\def\bcs{\begin{cases}}
\def\ecs{\end{cases}}
\def\bea{\begin{eqnarray}}
\def\eea{\end{eqnarray}}
\newcommand{\Aref}[1]{Appendix}%

\newcommand{\opunit}{\textrm{1}\kern-0.22em\textrm{l}}

\graphicspath{
	{./Images_supplementary/}
	{./Images_supplementary/ofv_is/}
	{./Images_supplementary/PD/}
	{./Images_supplementary/t_crystal/}
	{./Images_supplementary/fp/}
    {./Images_supplementary/MF_num_solving/}
}

\begin{document}


\title{Disorder induced time crystal in athermal random field Ising model with non-reciprocal interactions}

\author{{\normalsize{}Aldrin B E$^{1, 2}$}
{\normalsize{}}}
\email{aldrin.be@niser.ac.in}

\author{{\normalsize{}Sumedha$^{1, 2}$}
{\normalsize{}}}
\email{sumedha@niser.ac.in}

\affiliation{\noindent $^{1}$School of Physical Sciences, National Institute of Science Education and Research, Bhubaneswar, P.O. Jatni, 752050, India}

\affiliation{\noindent $^{2}$Homi Bhabha National Institute, Training School Complex, Anushakti Nagar 400094, India}


\begin{abstract}
A two species random field Ising model with non-reciprocal interactions between the species is studied  using the greedy Glauber dynamics. By solving the dynamics exactly on a complete graph, we obtain the phase diagram of the model as a function of the non-reciprocal interaction ($K$) and the variance ($\sigma$) of the quenched random field distribution. The model exhibits a rich phase diagram with the presence of a chaotic time-oscillatory phase for intermediate values of $K$ and $\sigma$. The chaotic phase has stable time oscillations  along with the autocorrelation time that diverges with system size on a complete graph and also in three dimensions. We find that the random field disorder along with non-reciprocal interaction alone can produce a time crystal without an external driving. In two dimensions the autocorrelation time does not increase with the system size and the time crystal phase is absent.
\end{abstract}

\maketitle

In physics, the traditional approach to many-body systems relies on defining a Hamiltonian. The dynamics of these systems typically 
exhibit time-reversal symmetry, which gives rise to specific conservation laws. However, this symmetry can be broken in various ways, most popularly by introducing non-reciprocal interactions \cite{alston, suchanek, fruchart, shi, mohite, fruchart2, hanai, morrell, avni, weiderpass, zhang, sethi}. Although the study of non-reciprocity is not new \cite{krichevtsov,roman,muthukumar,dumelow,godreche,kravtsov}, it has recently garnered significant attention due to its relevance to active matter \cite{dinelli,duan,kreienkamp,harraq,martin}, biological predator-prey dynamics \cite{lipowski,tauber,sarrola, antal}, micro-robotics \cite{brandenbourger}, photonics \cite{sounas,yang}, and open quantum systems \cite{nadolny,lau,hanai1}.

Breaking time-reversal symmetry inherently yields an out-of-equilibrium system. Recently, a novel phase of matter a ``time crystal", which sustains temporal oscillations with stable limit cycle, has been discovered. It results from the spontaneous breaking of time translation symmetry. Initial time crystals, created in 2016, required a periodic driving force to maintain synchronized oscillations and strong disorder (like in many-body localization) to break ergodicity and avoid thermalisation \cite{autti,giergiel,yao,khemani,choi,wilczek,daviet, yousefjani}. In contrast, non-reciprocal interactions can introduce frustration that pushes the system into a chaotic state without the need for external driving \cite{hanai,martin,tavakol}. In our study, we combine non-reciprocal interactions with quenched disorder to investigate the possibility of a time crystal phase in classical, out-of-equilibrium disordered systems.

Several models featuring discrete and continuous spins with non-reciprocal interactions have emerged recently. To model spin updates, some studies develop the Langevin approach \cite{fruchart, hanai, duan, kreienkamp, guislain, dinelli, lorenzana, stariolo, zhang}. Conversely, others employ Glauber dynamics\cite{avni, blom, weiderpass, garces}, a framework built for studying equilibrium steady states via single spin-flip updates \cite{shi,mohite}. 

In this paper, we study athermal dynamics of many-body spin systems in the presence of quenched Gaussian random fields, where the variance of the Gaussian random field provides the stochasticity. The dynamics is the single spin-flip greedy Glauber dynamics (GGD) where only the energy lowering flips are allowed (a gradient descent dynamics that aim to find the local minima).This allows us to model and study non-reciprocal interaction in Monte Carlo simulations without the use of Boltzmann weight. In GGD, if $\delta E$ is the change in energy involved in flipping a spin, then a randomly chosen spin $s_i(t)$ flips  with probability $W(\delta  E)$ given by 
\begin{align}
	W (\delta E) =
	\left\{
	\begin{array}{l}
		1~~~~\delta E<0 \\
		 \frac{1}{2}~~~\delta E = 0 \\
		 0 ~~~~ \delta E > 0
	\end{array}
   \right.
	\label{eq:1}
    \end{align}
 
 The GGD steady state for a Random field Ising model (RFIM) on a complete graph  with random field drawn from a Gaussian distribution $p(h)=\frac{1}{\sqrt{2 \pi \sigma^2}}  \exp(-h^2/2 \sigma^2)$ results in a self-consistent equation: 
 $m =\text{erf}(m/\sqrt{2}\sigma)$  for magnetisation which shows a continuous  non-equilibrium steady state transition at $\sigma_c=\sqrt{2/\pi}$. 
 
We define a two species non-reciprocal model using the Ising spins with quenched random field disorder. This is a disorder version of the non-reciprocal model introduced recently \cite{avni}. The model considers two species ($A$ and $B$) on each lattice site. Each  species exists in two states that are modeled by Ising spins $s_{iA}$  and $s_{iB}$  at each site $i$ and can take two possible values $\pm 1$.  The two  species interact ferromagnetically with the neighbors of their own species. There is also an inter species interaction on the same site : species $A$ interacts ferro-magnetically with the species $B$ and species $B$ interacts anti-ferromagnetically with $A$. The quenched random field $h_i$  at each site taken from the Gaussian distribution $p(h,\sigma)$, is same for both species $A$ and $B$. The greedy energy of species $A$ and $B$ is then given by 
\begin{align}
    E_{g(A)} &= -J_A \sum_{<ij>} s_{iA} s_{jA}-K \sum_i s_{iA} s_{iB}-\sum_i h_i s_{iA}\\
    E_{g(B)} &=-J_B \sum_{<ij>} s_{iB} s_{jB}+K \sum_i s_{iA} s_{iB}-\sum_i h_i s_{iB}
\end{align} 
here $J_A$ and $J_B$ are the ferromagnetic coupling between the neighbors of the same species and $K$ is the interaction strength between $A$ and $B$ species at the same site. 

For the two species model considered here, a randomly picked spin of species $\alpha(A/B)$ changes energy by an amount $\delta E_{i\alpha}=2s_{i\alpha} l_{i\alpha}$ when flipped. Here $l_{i\alpha}$ are the local fields, given by
\begin{eqnarray}
    l_{i\alpha} &=& J_{\alpha} \sum_{j\epsilon nn \text{ of } i} s_{j \alpha}+C_{i\alpha}+ h_i
    \label{eq:4}
\end{eqnarray} 
where $C_{iA}=Ks_{iB}$ and $C_{iB}=-K s_{iA}$.
As a result at time $t+1$ the updated spins under GGD are given by 
\begin{eqnarray}
    s_{i\alpha}(t+1) = sgn(l_{i\alpha}(t))
    \label{eq:5}
\end{eqnarray}
Note that GGD is different from the usual finite temperature Glauber dynamics \cite{glauber} where the rates are chosen to ensure the Boltzmann distribution in the steady state. In GGD, the dynamics is determined by the variance of the random field distribution. 

We first study the model on a complete graph by obtaining the steady state values of the order parameters $m_A$ and $m_B$ exactly. The  Eqs 1 and 2 on a complete graph become
\begin{eqnarray}
    E_{g(\alpha)} &=& -\frac{J}{2 N}( \sum_{i} s_{i\alpha})^2 -\sum_i s_{i\alpha} C_{i\alpha}-\sum_i h_i s_{i\alpha}
    \label{eq:6}
\end{eqnarray} 
where $J_{\alpha}=J/N$. The steady state of the dynamics can be solved exactly on a complete graph (see appendix). We find that making a mean field assumption for inter species interaction term, namely $K s_{iB}= K m_B$ and $Ks_{iA}=K m_A$ $\forall i$  gives similar behavior. We  hence  make  this assumption here to calculate the GGD steady state.

Let $P_{\alpha}(+1)$ be the probability for a randomly chosen spin of species $\alpha$ to be up ($+1$). Then the average value of spins of species $\alpha$ is equal to $m_{\alpha} = 2 P_{\alpha}(+1)-1$. This gives self-consistent equations for $m_{\alpha}$ as:
\begin{align}
m_{\alpha} = 2 \int_{-x_{\alpha}}^{\infty} d h p(h,\sigma)-1 = \text{erf} \left(\frac{x_{\alpha}}{\sqrt{2} \sigma}\right) 
\label{eq:7}
\end{align}
where $x_A=J m_A+K m_B$ and $x_B=Jm_B-K m_A$.
The GGD reaches a fixed point of the above simultaneous equations in the steady state.

Interestingly, the dynamics does not always reach a fixed point and sometimes enters a limit cycle. This can be seen by considering the simpler case where we set $J=0$. The system reduces to a single site system with update rules: $s_{A} = sgn(K s_B+h)$ and $s_B=sgn(-Ks_A+h)$, where $h$ is a quenched i.i.d random variable from the distribution $p(h,\sigma)$. In this case if $K$ is large enough, the spins keep on flipping as spin $s_A$ keep chasing $s_B$ and tries to make $s_A=s_B$. On the other hand, $B$ tries to escape $A$ by preferring a state with $s_B=-s_A$. This gives rise to  perfectly synchronized single  site oscillations in the system for large $K$. In this paper we show that in the presence of $K$ for $J \neq 0$, the collective oscillations build up in the system which are sustained in the system for a very long time. The presence of a chaotic state with time oscillations depends on the interplay of $K$ and $\sigma$.

The $E_{gA}$ and $E_{gB}$ have the form similar to the Hamiltonian of the RFIM in an external field. The spins of the two species, though coupled, can be seen as moving in a free energy landscape of the RFIM with an external field proportional to $K m_{A/B}$. The RFIM in an external field at zero temperature can be solved exactly on a complete graph with Gaussian distribution of the random fields (see \cite{sumedha} for method). The disorder averaged free energy potential of species $\alpha$ is 
\begin{align}
f_{\alpha}  &= \frac{J}{2} m_{\alpha}^2-x_{\alpha} \text{erf}\left(\frac{x_{\alpha}}{\sqrt{2} \sigma}\right)- 
\sigma \sqrt{\frac{2}{\pi}} \exp\left(-\frac{x_{\alpha}^2}{2 \sigma^2}\right).
\label{eq:8}
\end{align}
\begin{figure}[t]
	\centering
	\includegraphics[width=0.45\textwidth]{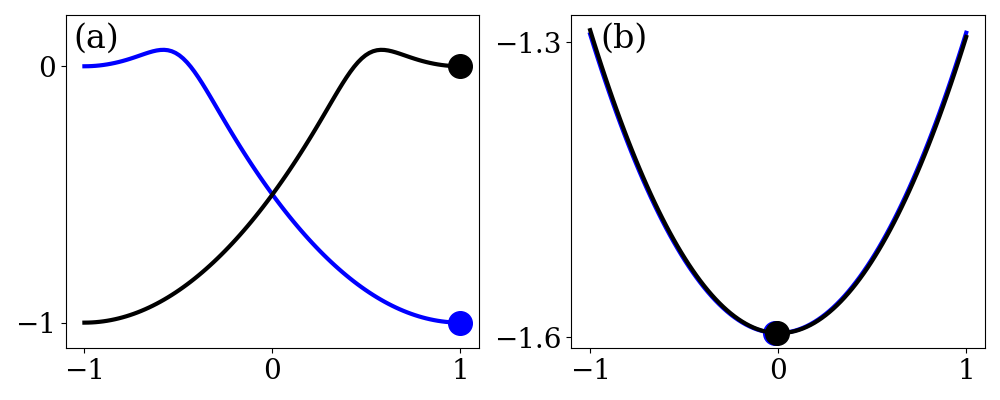}
	\caption{The free energy functionals $f_A$ (blue) and $f_B$(black) are plotted as a function of $m_A, m_B$ respectively in the steady state of a single run. The $m_A$(blue dot) and $m_B$(black dot) steady state values are also plotted: a) $K = 0.5, \sigma = 0.1$ where an ordered steady state with $m_A, m_B = +1$ is reached; b) $K = 0.5, \sigma = 2$ with disordered state with $m_A=m_B=0$.}
    \label{fig:1}
\end{figure}
\begin{figure}[t]
	\centering
	\includegraphics[width=0.47\textwidth]{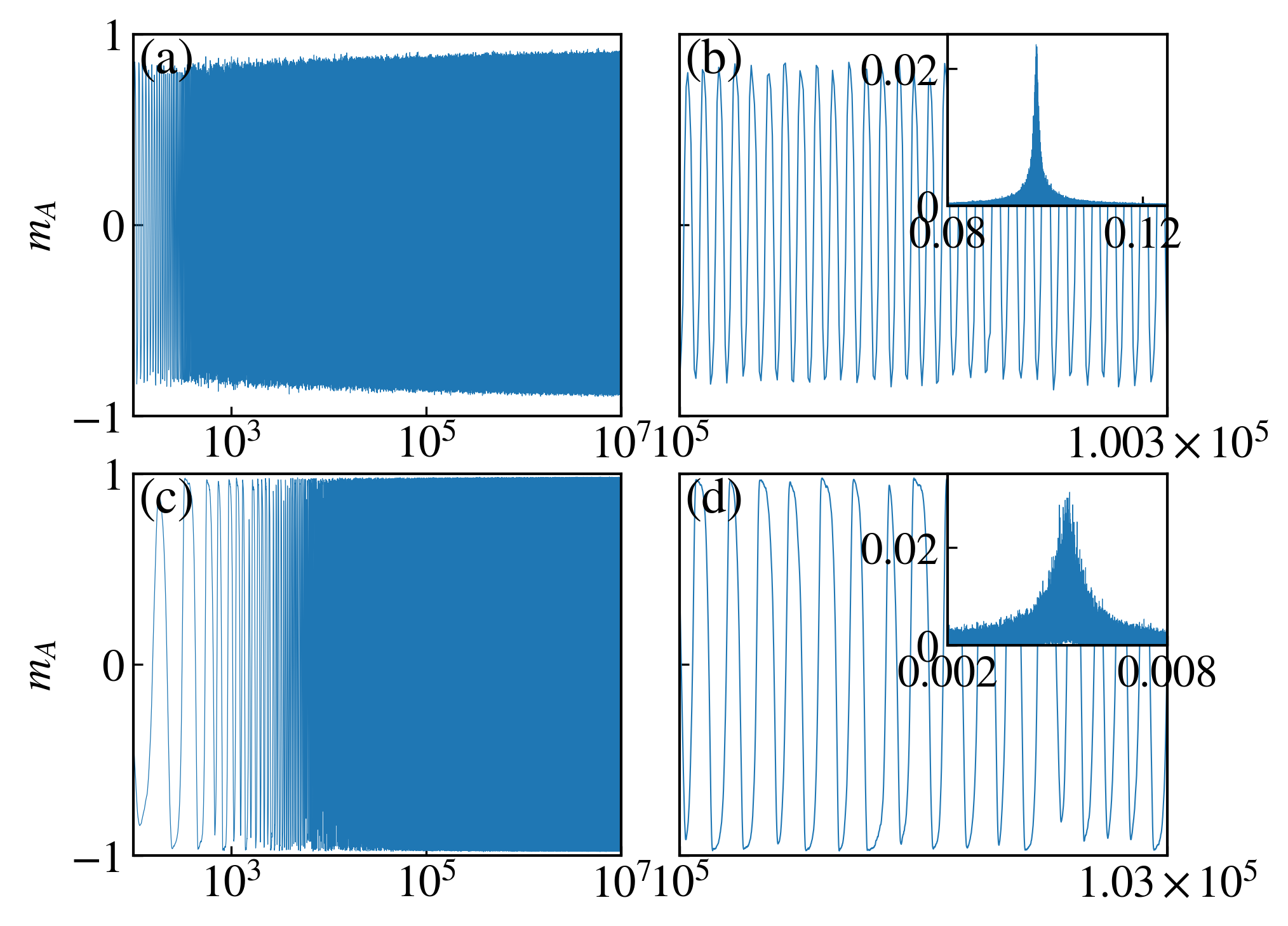}
	\caption{Evolution of $m_A$ with time is plotted for a randomly picked realization for: a,b) Complete graph with  $K = 0.5,\sigma = 0.4$ for $N = 1000$. c,d) Cubic lattice with $K = 0.3,\sigma = 0.3$ for $L = 30$. In the insets the Fourier transform $\left(\mathcal{F}(m_A)\right)$ of $m_A(t)$ are plotted as a function of frequency.}
	\label{fig:2}
\end{figure}
\begin{figure}[t]
	\centering
\includegraphics[width=0.44\textwidth]{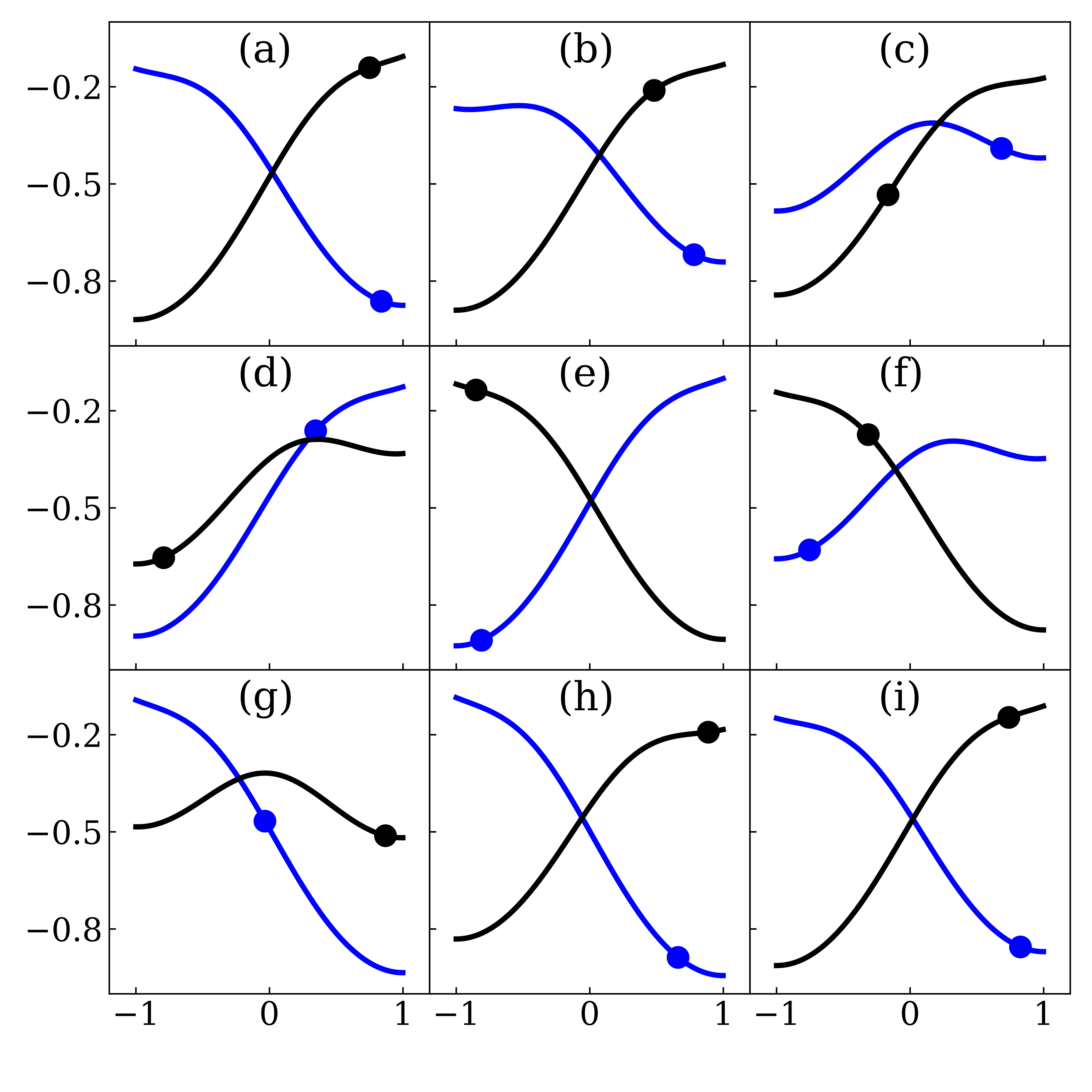}
	\caption{Time evolution of $m_A$ and $m_B$ plotted on the dynamic energy landscape from Eq. \ref{eq:8} for $J = 1; K = 0.5; \sigma = 0.4$ on a complete graph of size $N = 1000$. The subfigures (a) to (i) correspond to the states reached after $501$, $502$, $503$, $504$, $506$, $508$, $510$, $511$, \text{and }$512$ Monte Carlo runs respectively. These states keep on repeating themselves.}
\label{fig:3}
\end{figure}
\begin{figure}[t]
	\centering
	\includegraphics[width=0.47\textwidth]{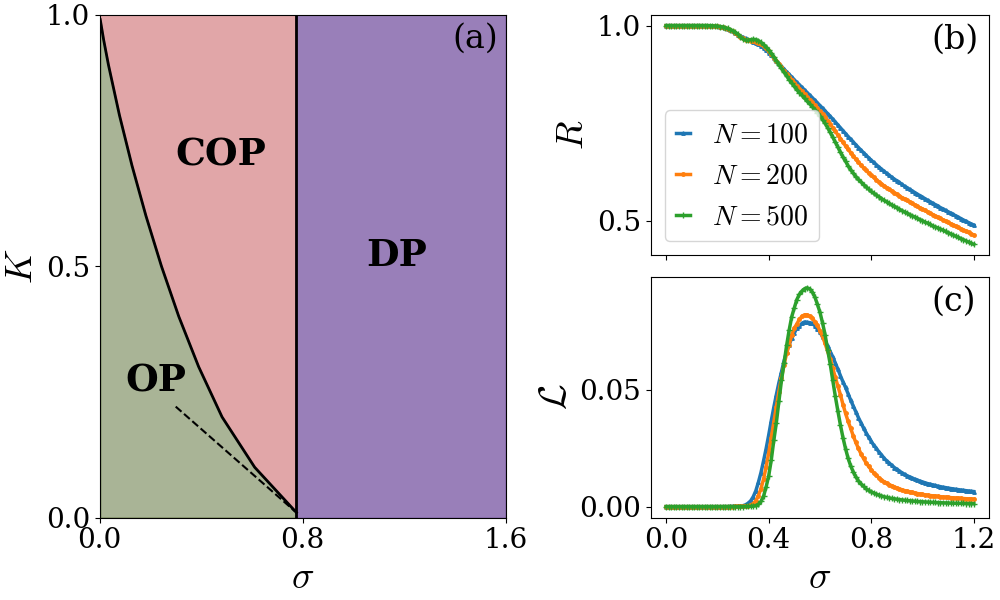}
	\caption{ The phase diagram in $K-\sigma$ plane on a complete graph is plotted in (a) with the ordered  phase (OP), chaotic ordered  phase (COP) and disordered  phase (DP). The solid lines mark the lines of transition between the two phases from the full numerical solution of Eq. \ref{eq:9}. The dashed line is from the solution of the cubic expansion of the dynamical equations ($K = (1-\sqrt{\pi/2}\ \sigma)/(2\sqrt{2})$). In (b) and (c) the order parameters $R$ and $\mathcal{L}$, obtained via Monte Carlo simulations, are plotted for $K = 0.4$ as a function of $\sigma$, averaged over the initial states $m_A=m_B=1$ and $m_A=-m_B=1$, over $5\times10^4$, $3\times10^4$ and $10^4$ realizations for system sizes $N = 100, 200,$ and $500$ respectively.}
	\label{fig:4}
\end{figure}
\begin{figure*}[t]
	\centering
	\includegraphics[width=1.0\textwidth]{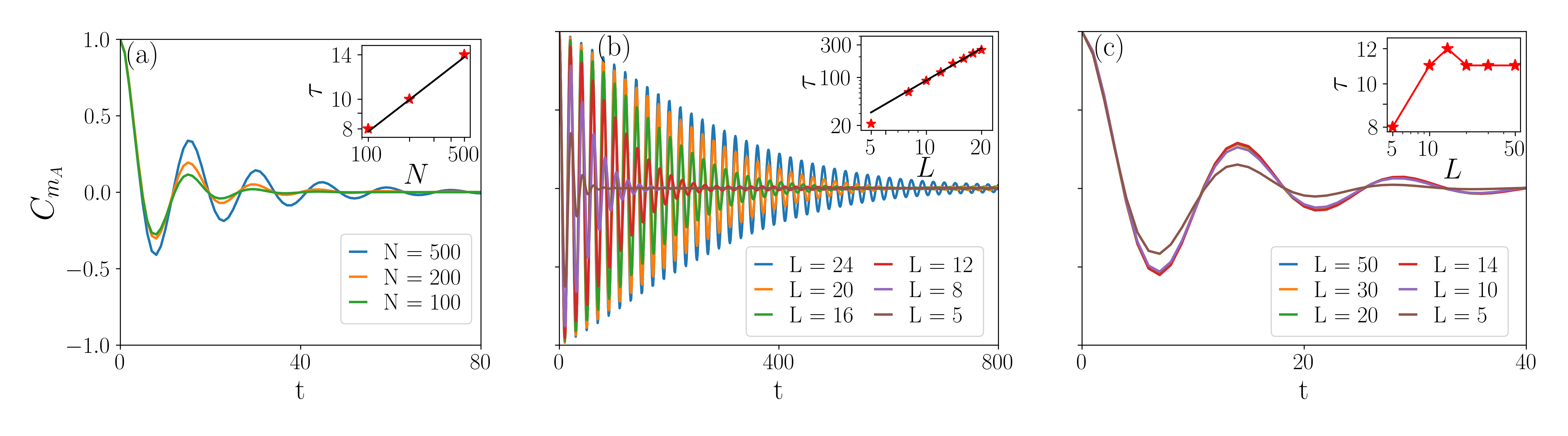}
	\caption{The time-autocorrelation function ($C_{m_A}(t)$) for $m_A$ for (a) complete graph, (b) cubic lattice  and (c) square lattice. The autocorrelation time ($\tau$)  shown in inset  is calculated by finding the intersection of the upper envelope of $C_{m_A}(t)$ with $e^{-1}$. $\tau$ is plotted as a function of system size in log-log scale in the insets. For the complete graph and the cubic lattice, the values of $\tau$ obtained for each system size was fitted with the function $aL^{b}$ and we obtain $a = 1.56 \pm 0.11; b = 0.35 \pm 0.01$ and $a = 2.55 \pm 0.39; b = 1.55 \pm 0.06$ respectively. For the case of square lattice, $\tau$ does not have system size dependence. }
	\label{fig:5}
\end{figure*}

The equations $\partial f_{\alpha}/\partial m_{\alpha} =0$ gives the fixed point equations. These are exactly the same as Eqs. \ref{eq:7}. The time evolution of $m_{\alpha}$  is given by the equations $dm_{\alpha}/dt = -\partial f_{\alpha}/\partial m_{\alpha}$ \cite{halperin}. This results in the equations:
\begin{align}
 \delta \frac{d m_{\alpha}(t)}{dt} &= -m_{\alpha}+\text{erf} \left(\frac{x_{\alpha}}{\sqrt{2} \sigma}\right)  \label{eq:9}
\end{align}
here $\delta$ is the time of single Monte Carlo step, set to $ 1$ for the rest of the paper. These equations are the dynamical equations for the GGD on a complete graph in the presence of random field disorder. Hence, even though the Glauber dynamics didn't had an underlying free energy potential, $f_A$ and $f_B$ act as the free energy potential for the dynamics defining the underlying dynamic energy landscape on which the order parameter of two species move under the GGD. The fixed points of the dynamics are the values of $m_A$ and $m_B$ such that both A and B subsystems are respectively in the local minima of $f_A$ and $f_B$. In Fig. \ref{fig:1} we have plotted the steady state values of $m_A$ and $m_B$ over the functions $f_A$ and $f_B$. For large $\sigma$, as expected $m_A=m_B=0$ is the steady state and for low $\sigma$, $m_A$ and $m_B$ both have reached a local minima in the steady state with $m_A\approx m_B \approx 1$.
At intermediate values of $\sigma$ the system fails to reach a steady state. In this case one finds that $m_A$ and $m_B$ never reach minima of $f_A$ and $f_B$ simultaneously. The system goes into a time oscillatory phase where $m_A$ and $m_B$ exhibit oscillations that do not decay with time. These are shown in Fig \ref{fig:2} (a) and (b) for complete graph and in Fig. \ref{fig:2} (c) and (d) for cubic lattice, obtained by simulating the model using GGD till $10^7$ Monte Carlo steps. One time step corresponds to a sweep over both A and B species over the entire lattice. The oscillations are robust and show no sign of dying out. Also, the Fourier  transform shows a peak that grows  with system size. In the case of a complete graph the corresponding position of the $m_A$ and $m_B$ on $f_A$ and $f_B$ are shown respectively in Fig. \ref{fig:3}. 

The presence of sustained time oscillations suggests the presence of a time ordered chaotic phase in the system. The phase diagram under Glauber dynamics can be obtained exactly on a complete graph via studying Eqs. \ref{eq:9}. Since in our model the noise can be tuned independently of the couplings by tuning $\sigma$, we obtain phase diagram between $K$ and $\sigma$ for a fixed $J=1$ to understand the interplay of non-reciprocal interaction and noise. The phase diagram is shown in Fig. \ref{fig:4} (see appendix for details). The system transitions from a disordered phase to a chaotic ordered phase  at $\sigma =\sqrt{2/\pi}$  via  Hopf bifurcation where the order parameters $m_A$ and $m_B$ are of the form $r e^{-i \theta}$. This phase exhibits large amplitude oscillations due to collective motion of spins, a time crystal. It is a continuous time quasi crystal phase as the presence of disorder gives rise to robust non-repeating pattern in oscillations. Another transition occurs between the chaotic ordered phase and ordered phase. This boundary can be obtained by  solving the dynamical equations in polar coordinates (see appendix). We find that the line separating the chaotic ordered phase from ordered phase matches with the line that separates the region with only $(0,0)$  fixed point from the region with non-zero fixed  points in  $K-\sigma$ plane. For $K>1$ there are no non-zero fixed points to the Eqs. \ref{eq:7} and the system exists in a chaotic state with small amplitude(single site) oscillations for all $\sigma$.

We can also obtain a phase diagram in $\widetilde{J}-\widetilde{K}$ plane, where $\widetilde{J}=J/\sigma;\widetilde{K}=K/\sigma$ as was done for the pure model with $\sigma$ being replaced by temperature ($T$) in the earlier study \cite{avni}. We obtain similar phase diagram in $\widetilde{J}-\widetilde{K}$ plane. However, from such a phase diagram, the interplay of non-reciprocity and disorder cannot be understood straightforwardly.

We perform Monte-Carlo simulations of the model and study the following two order parameters a) An  order parameter to separate the ordered phase from the disordered phase given by $R = \bigg\langle \sqrt{\frac{m_A^2+m_B^2}{2}} \bigg\rangle_{t,Q}$.  The average is performed over $t$ Monte Carlo steps and $Q$ disorder realizations and b) phase space angular momentum $\mathcal{L} =\big\langle l\big\rangle_{t,Q}$ with $l(t) = m_B(t)\partial_t m_A(t) - m_A(t)\partial_t m_B(t)= m_A(t)m_B(t-1)-m_B(t)m_A(t-1)$ for a given realization. The phase space angular momentum becomes non-zero in the presence of long range collective oscillations. For a complete graph, the study of these order parameters verify the phase diagrams (see Fig. \ref{fig:4} and appendix).  A non-zero value of $\mathcal{L}$ along with snap shots shown in Fig. \ref{fig:2}, confirms the presence of a time crystal in the chaotic ordered phase. 

We  studied also the autocorrelation function of magnetization $m_{\alpha}$ and the corresponding autocorrelation time $\tau$. It is plotted in Fig. \ref{fig:5}. For a complete graph, it increases with $N$, with a power law $\tau \sim N^{0.35\pm0.01}$. The plots for complete graph are obtained by averaging over $100$ realizations of random field, with $100$ time series per realization, each with $4000$ time steps. 

In order to see if the time ordered phase appear also in finite dimensions, we simulated the model on cubic and square lattices using GGD. In the case of cubic lattice behaviour was similar to complete graph, the intermediate values of disorder and non-reciprocal strength resulted in a time crystal phase. The snap shot of the oscillations is shown in Fig. \ref{fig:2} till $10^7$ Monte Carlo steps. The corresponding frequency plot shows a peak that gets sharper with system size (see appendix). The autocorrelation function also showed a growth with system size, with autocorrelation time scaling as $\tau \sim  L^{1.55 \pm 0.06}$ (see Fig. \ref{fig:5}). The $\tau$ dependence is lower than $L^3$ due to the fluctuations induced by random field which typically go as inverse of the square root of the system size. The simulations on cubic lattice are hard due to lack of self averaging and are sensitive to the number of realizations studied. This limits the value of $L$ that can be studied.  The plots of cubic lattice are averaged over $1000$ realizations, with $100$ time series per realization, each with $10^4$ time steps. In contrast, in two dimensions on a square lattice, the autocorrelation function has no dependence on the system size. The autocorrelation time $\tau$ tends to a constant with increasing system size. 

We find that system does not stabilize in the times that we could realistically study (upto $10^7$ Monte Carlo runs). Hence besides the dynamic exponent, related to $\tau$, we have not discussed the scaling of the other quantities.

{\it Discussion:} We introduced an athermal model with non-reciprocal interactions that is exactly solvable under Glauber dynamics. Because non-reciprocal interactions are inherently non-equilibrium in nature, standard frameworks designed for equilibrium steady states do not guarantee correspondence to steady state of a microscopic Langevin dynamics \cite{shi}. Rescaling coupling by $T$ does not make the dynamics temperature independent. By introducing stochasticity via random-field disorder, we provide an alternative approach to study non-reciprocity in non-equilibrium set up. As a result we could unravel the interplay of noise and non-reciprocal interaction. This framework is generalizable to any non-reciprocal model.

We have shown that disorder and non-reciprocal interactions are sufficient to generate a time crystal phase without external driving. This phase occurs at intermediate strengths of both parameters: weak or strong disorder fails to produce a chaotic ordered state, while large non-reciprocity reduces the system to an effectively single site system. On a complete graph, the existence of the time crystalline phase is established both analytically and numerically. In three dimensions, strong evidence for this phase emerges through long time oscillations and a system-size-dependent increase in autocorrelation time. While current numerical results strongly indicate a time quasi-crystal, larger system size study is required in three dimensions. With increasing system size, one requires averaging over larger number of disorder realizations and a study with more Monte Carlo steps in each run. These considerations limits the system size that can be studied in simulations. In two dimensions, a time crystal phase is ruled out.

\appendix

\section{Exact Glauber steady-state solution on a complete graph}
The local fields for greedy Glauber dynamics at a site $i$ for a spin of species $A$ and $B$ are:
\begin{equation}
    l_{iA} = \frac{J}{N}\sum_{i \neq j} s_{jA} + K s_{iB} + h_i
    \label{Eq:l_A}
\end{equation}
\begin{equation}
    l_{iB} = \frac{J}{N}\sum_{i \neq j} s_{jB} - K s_{iA} + h_i
    \label{Eq:l_B}
\end{equation}
The $h_i$ are sampled from a Gaussian distribution of mean zero and standard deviation $\sigma$, $J$ is the interaction strength between neighbouring sites of the same species, and $K$ is the non-reciprocal interaction between the two spins of different species on the same lattice site.\\
We define
$\mathcal{P}^\pm_{A/B} = \text{Prob}\big(l_{iA/B} > 0\ \big| \ s_{iB/A} = \pm 1\big)$ and $p^\pm_{A/B}= \text{Prob}\big(s_{iA/B} = \pm 1\big)$. Since it is a complete graph, all sites $i$ are equivalent. Hence, we remove the index $i$. The probabilities $p^+_A$ and $p^+_B$ are respectively
\begin{equation}p^+_A = \mathcal{P}^+_A p^+_B + \mathcal{P}^-_A (1-p^+_{B})
	\label{eq:P_A}
\end{equation}

\begin{equation}p^+_B = \mathcal{P}^+_B p^+_A + \mathcal{P}^-_B (1-p^+_{A})
	\label{eq:P_B}
\end{equation} 
where
$\mathcal{P}^\pm_{A} =\frac{1}{2} \Big(1+\text{erf}\big(\frac{Jm_A\pm K}{\sqrt{2}\sigma}\big)\Big)	$, 
$ \mathcal{P}^\pm_{B} = \frac{1}{2} \Big(1+\text{erf}\big(\frac{Jm_B\mp K}{\sqrt{2}\sigma}\big)\Big)$.\\
From Eqs. \ref{eq:P_A} and \ref{eq:P_B} and using $m_\alpha=2p^+_{\alpha} - 1 $ for a given species $\alpha$, we get
\begin{equation}
	m_A = \frac{\Big(\mathcal{P}^+_{A} - \mathcal{P}^-_{A}\Big)\Big(\mathcal{P}^+_{B} + \mathcal{P}^-_{B}\Big)+2\ \mathcal{P}^-_{A}- 1}{1-\Big(\mathcal{P}^+_{A}-\mathcal{P}^-_{A}\Big)\Big(\mathcal{P}^+_{B}-\mathcal{P}^-_{B}\Big)}
	\label{eq:m_A_final}
\end{equation}

\begin{equation}
	m_B = \frac{\Big(\mathcal{P}^+_{A} + \mathcal{P}^-_{A}\Big)\Big(\mathcal{P}^+_{B} - \mathcal{P}^-_{B}\Big)+2\ \mathcal{P}^-_{B}- 1}{1-\Big(\mathcal{P}^+_{A}-\mathcal{P}^-_{A}\Big)\Big(\mathcal{P}^+_{B}-\mathcal{P}^-_{B}\Big)}
	\label{eq:m_B_final}
\end{equation}

Substituting $\mathcal{P}^+_A,\ \mathcal{P}^-_A,\ \mathcal{P}^+_B,\ \mathcal{P}^-_B$,
\begin{align}
	m_A = &\notag \frac{1}{2D}\bigg[\text{erf}\bigg(\frac{Jm_A+K}{\sqrt{2}\sigma}\bigg)+\text{erf}\bigg(\frac{Jm_A-K}{\sqrt{2}\sigma}\bigg)\\ &+\frac{1}{2}\bigg(\text{erf}\bigg(\frac{Jm_A+K}{\sqrt{2}\sigma}\bigg)-\text{erf}\bigg(\frac{Jm_A-K}{\sqrt{2}\sigma}\bigg) \bigg)\notag\\ &\bigg( \text{erf}\bigg(\frac{Jm_B+K}{\sqrt{2}\sigma}\bigg)+\text{erf}\bigg(\frac{Jm_B-K}{\sqrt{2}\sigma}\bigg) \bigg)\bigg]\label{eq:exact_mA}
\end{align}

\begin{align}
	m_B = &\notag \frac{1}{2D}\bigg[\text{erf}\bigg(\frac{Jm_B+K}{\sqrt{2}\sigma}\bigg)+\text{erf}\bigg(\frac{Jm_B-K}{\sqrt{2}\sigma}\bigg)\\ &+\frac{1}{2}\bigg(\text{erf}\bigg(\frac{Jm_B+K}{\sqrt{2}\sigma}\bigg)-\text{erf}\bigg(\frac{Jm_B-K}{\sqrt{2}\sigma}\bigg) \bigg)\notag\\ &\bigg( \text{erf}\bigg(\frac{Jm_A+K}{\sqrt{2}\sigma}\bigg)+\text{erf}\bigg(\frac{Jm_A-K}{\sqrt{2}\sigma}\bigg) \bigg)\bigg]\label{eq:exact_mB}
\end{align}

where \begin{align*}
	D = 1-\frac{1}{4}&\bigg(\text{erf}\bigg(\frac{Jm_A+K}{\sqrt{2}\sigma}\bigg)-\text{erf}\bigg(\frac{Jm_A-K}{\sqrt{2}\sigma}\bigg)\bigg)\\ &\bigg(\text{erf}\bigg(\frac{Jm_B+K}{\sqrt{2}\sigma}\bigg)-\text{erf}\bigg(\frac{Jm_B-K}{\sqrt{2}\sigma}\bigg)\bigg) 
\end{align*}

\begin{figure}
	\centering
	\includegraphics[width=0.9\linewidth]{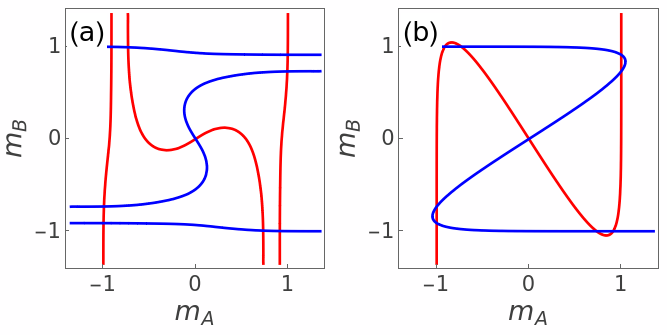}
	\caption{Contour plots obtained from the exact Glauber steady-state equations (Eqs. \ref{eq:exact_mA} (in red) and \ref{eq:exact_mB} (in blue)) in subfigure (a) and from the mean-field fixed point equations (Eqs. \ref{Eq:MF_fp_m_A} (in red) and \ref{Eq:MF_fp_m_B} (in blue)) in subfigure (b) for the case of $J = 1;K = 0.4;\sigma = 0.3$. The intersection of the two contour lines correspond to the fixed point solutions. Even though the contours are not exactly the same, the fixed points are same in both the cases. }
	\label{fig:fp_J1_K0.4_R0.3}
\end{figure}

We can solve for the $m_A$ and $m_B$ numerically, by using the above self-consistent equations. Fig. \ref{fig:fp_J1_K0.4_R0.3}(a) shows fixed point of the equation obtained numerically for Eqs. \ref{eq:exact_mA} and \ref{eq:exact_mB}. \\We can simplify the above expressions by making the mean-field assumption: replacing $Ks_{iA/B}$ by $Km_{A/B}$ in the local fields $l_{iB/A}$ (Eqs. \ref{Eq:l_A}, \ref{Eq:l_B}). In that case self-consistent equations are

\begin{equation}
m_A = \text{\text{erf}}\bigg(\frac{Jm_A + Km_B} {\sqrt{2}\sigma} \bigg)
\label{Eq:MF_fp_m_A}
\end{equation}

\begin{equation}
m_B = \text{\text{erf}}\bigg(\frac{Jm_B-Km_A}{\sqrt{2}\sigma}\bigg)
\label{Eq:MF_fp_m_B}
\end{equation}

\begin{figure*}
	\centering
	\includegraphics[width=0.8\linewidth]{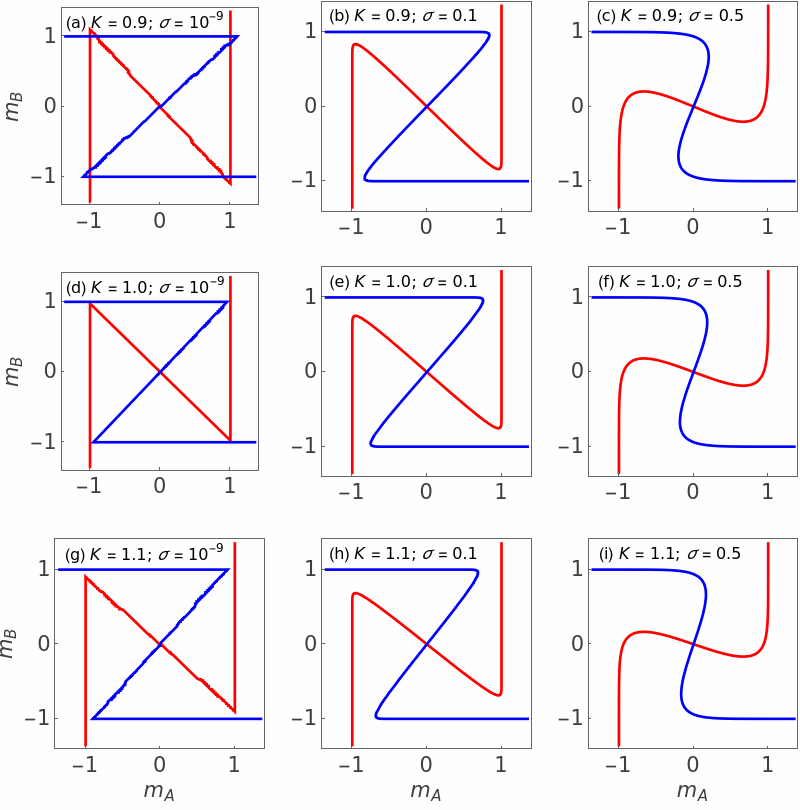}
	\caption{Fixed point solutions to Eqs. \ref{Eq:MF_fp_m_A} (in red) and \ref{Eq:MF_fp_m_B} (in blue) are plotted for different values of $K$ and $\sigma$ for $J = 1$. We see that for $K\geq 1$ the non-zero solution ceases to exist even for low values of $\sigma$.}
	\label{fig:fp_approx_J1}
\end{figure*}

 We checked for various sets of parameters and found that the fixed points from the exact (Eqs. \ref{eq:exact_mA}, \ref{eq:exact_mB}) and the mean-field (Eqs. \ref{Eq:MF_fp_m_A}, \ref{Eq:MF_fp_m_B}) equations have similar nature and are roughly at the same value of $m_A$ and $m_B$ (Fig. \ref{fig:fp_J1_K0.4_R0.3}).\\
 We also found that for $K>J$, $m_A = m_B = 0$ is the only fixed point in both cases. This can be seen via contradiction from Eqs. \ref{Eq:MF_fp_m_A} and \ref{Eq:MF_fp_m_B}. In Fig. \ref{fig:fp_approx_J1} we have plotted their fixed points. Hence for $K > J$ the system cannot have any ordered state. 

\section{Solution to the dynamic equations on a complete graph}

The calculations in the following subsection follow a procedure similar to Avni et al. in \cite{avni2}.
The dynamic equations on a complete graph under mean-field assumption are:
\begin{equation}
	\partial_t m_A = -m_A + \text{\text{erf}}\bigg(\frac{Jm_A + Km_B}{\sqrt{2}\sigma}\bigg)
	\label{Eq:dyna_m_A}
\end{equation}
\begin{equation}
	\partial_t m_B = - m_B + \text{\text{erf}}\bigg(\frac{Jm_B-Km_A}{\sqrt{2}\sigma}\bigg)
	\label{Eq:dyna_m_B}
\end{equation}
We define
$$F_A(m_A, m_B) = -m_A + \text{\text{erf}}\bigg(\frac{Jm_A + Km_B}{\sqrt{2}\sigma}\bigg)$$
$$F_B(m_A, m_B) = - m_B + \text{\text{erf}}\bigg(\frac{Jm_B-Km_A}{\sqrt{2}\sigma}\bigg)$$

Linearizing near a fixed point $(m^*_A, m^*_B)$, the Eqs. \ref{Eq:dyna_m_A} and \ref{Eq:dyna_m_B} can be written as:

\begin{align*}
\begin{bmatrix}
	\partial_t \delta m_A \\
	\partial_t \delta m_B
\end{bmatrix}
=\begin{bmatrix}
	\frac{\partial F_A}{\partial m_A} & \frac{\partial F_A}{\partial m_B} \\
	\frac{\partial F_B}{\partial m_A} & \frac{\partial F_B}{\partial m_B}
\end{bmatrix}&\Bigg|_{(m^*_A, m^*_B)}
\begin{bmatrix}
	\delta m_A \\
	\delta m_B
\end{bmatrix}\\= 
\mathcal{J}\big|_{(m^*_A, m^*_B)}
\begin{bmatrix}
	\delta m_A \\
	\delta m_B
\end{bmatrix}
\end{align*}

We define $X = \frac{Jm_A^* + Km_B^*}{\sqrt{2}\sigma}$, $Y = \frac{Jm_B^* - Km_A^*}{\sqrt{2}\sigma}$ and using $\frac{d}{dx}\text{\text{erf}}(x) = \frac{2}{\sqrt{\pi}}e^{-x^2}$, we get the Jacobian
$$\mathcal{J} = \begin{bmatrix}
	-1+\sqrt{\frac{2}{\pi}}\big(\frac{J}{\sigma}\big)e^{-X^2} & \sqrt{\frac{2}{\pi}}\big(\frac{K}{\sigma}\big)e^{-X^2}  \\
	-\sqrt{\frac{2}{\pi}}\big(\frac{K}{\sigma}\big)e^{-Y^2}  & -1+\sqrt{\frac{2}{\pi}}\big(\frac{J}{\sigma}\big)e^{-Y^2}
\end{bmatrix}$$
In order to find the flow near $(m^*_A, m^*_B) = (0, 0)$, we evaluate the Jacobian at that fixed point. It is
$$\mathcal{J} = \begin{bmatrix}
	-1+\sqrt{\frac{2}{\pi}}\big(\frac{J}{\sigma}\big) & \sqrt{\frac{2}{\pi}}\big(\frac{K}{\sigma}\big)  \\
	-\sqrt{\frac{2}{\pi}}\big(\frac{K}{\sigma}\big)   & -1+\sqrt{\frac{2}{\pi}}\big(\frac{J}{\sigma}\big)
\end{bmatrix}$$
The characteristic equation $|\mathcal{J}-\lambda I| = 0$ gives the Eigenvalues
\begin{equation}
\lambda = \sqrt{\frac{2}{\pi}} \bigg(\frac{J}{\sigma}\bigg) - 1 \pm i \sqrt{\frac{2}{\pi}} \bigg(\frac{K}{\sigma}\bigg)
\label{Eq:eigenvalue}
\end{equation}
For the case of $K = 0$, there are degenerate Eigenvalues $\lambda = \sqrt{\frac{2}{\pi}} \big(\frac{J}{\sigma}\big) - 1$. Thus, $(m^*_A, m^*_B) = (0,0)$ is a stable fixed point for $\sqrt{\frac{2}{\pi}} \big(\frac{J}{\sigma}\big) - 1 < 0$ or $\sigma>\sqrt{\frac{2}{\pi}}J$.\\ For non-zero $K$, for $\sigma > \sqrt{\frac{2}{\pi}} J$, there are oscillatory solutions. To study these solutions, we expand $\text{erf}(x)$ for small $x$. For small $x$, $\text{\text{erf}}(x) \approx \frac{2}{\sqrt{\pi}}\bigg(x - \frac{x^3}{3}\bigg)$. Assuming $m_A, m_B$ to be small, one can write Eqs. \ref{Eq:dyna_m_A} and \ref{Eq:dyna_m_B} , keeping terms upto cubic order in $m_A$ and $m_B$, as follows:

\begin{align}
	\dot{m}_A =& \Bigg(b\bigg(\frac{J}{\sigma}\bigg)-1\Bigg)m_A + b\bigg(\frac{K}{\sigma}\bigg)m_B\notag\\&-\frac{1}{6}b \Bigg(\bigg(\frac{J}{\sigma}\bigg)m_A+\bigg(\frac{K}{\sigma}\bigg)m_B\Bigg)^3
	\label{Eq:dyna_approx_mA}
\end{align}

\begin{align}
	\dot{m}_B =& \Bigg(b\bigg(\frac{J}{\sigma}\bigg)-1\Bigg)m_B - b\bigg(\frac{K}{\sigma}\bigg)m_A\notag\\& -\frac{1}{6}b \Bigg(\bigg(\frac{J}{\sigma}\bigg)m_B-\bigg(\frac{K}{\sigma}\bigg)m_A\Bigg)^3
	\label{Eq:dyna_approx_mB}
\end{align}
where $b = \sqrt{\frac{2}{\pi}}$. For $K = 0$ the equations are 
$$
\dot{m}_A =m_A\Bigg(\Bigg(b\bigg(\frac{J}{\sigma}\bigg)-1\Bigg)-\frac{1}{6}b \bigg(\frac{J}{\sigma}\bigg)^3 m_A^2\Bigg) 
$$

$$
\dot{m}_B = m_B \Bigg(\Bigg(b\bigg(\frac{J}{\sigma}\bigg)-1\Bigg) -\frac{1}{6}b \bigg(\frac{J}{\sigma}\bigg)^3 m_B^2\Bigg)
$$
We solve the above equations for the fixed points setting $\dot{m}_A = \dot{m}_B = 0$.\\ 
For $\sigma>\sqrt{\frac{2}{\pi}}J$, the only real solution is $(m_A^*,m_B^*) = (0,0)$ which is a stable fixed point. \\
For $\sigma<\sqrt{\frac{2}{\pi}}J$, we have two stable fixed points $(m_A^*,m_B^*) = \left(\pm\sqrt{\frac{6\big(b\big(\frac{J}{\sigma}\big)-1\big)}{b\big(\frac{J}{\sigma}\big)^3}}, \pm\sqrt{\frac{6\big(b\big(\frac{J}{\sigma}\big)-1\big)}{b\big(\frac{J}{\sigma}\big)^3}}\right)$ and one unstable fixed point $(m_A^*,m_B^*) = (0,0)$.\\Thus, at $K = 0$, there is a Pitchfork bifurcation at $\sigma=\sqrt{\frac{2}{\pi}}J$.\\
For the case of $K > 0$, the Eigenvalues (Eq.  \ref{Eq:eigenvalue}) indicate that there is angular evolution of the system. Here, for $K>0$ there is a Hopf bifurcation at $\sigma=\sqrt{\frac{2}{\pi}}J$. We discuss the dynamical equations in polar co-ordinates to study this phase in the next subsection. We have noticed in the previous section that there are no non-trivial fixed point for $K >J=1$.

\subsection{Polar co-ordinates} In order to differentiate between an oscillatory phase and a steady state, it is beneficial to write the dynamical equations in terms of the polar co-ordinates. With $m_A = r \cos\theta$ and $m_B = r \sin\theta$, we write the dynamic equations (Eqs. \ref{Eq:dyna_m_A} and \ref{Eq:dyna_m_B}) in polar co-ordinates as follows:
\begin{equation}
	\partial_t r = -r +X\cos\theta+Y\sin\theta
	\label{partial_t_r}
\end{equation}
\begin{equation}
	r\partial_t \theta = Y\cos\theta-X\sin\theta
	\label{partial_t_theta}
\end{equation}
where $X = \text{\text{erf}}\big(a(J\cos\theta + K\sin\theta)\big)$, 
$Y = \text{\text{erf}}\big(a(J\sin\theta - K\cos\theta)\big)$, and $a = \frac{r}{\sqrt{2}\sigma}$. Under the approximation that $m_A$ and $m_B$ are small and keeping terms only upto the cubic order in $m_A \text{ and } m_B$ (from Eqs. \ref{Eq:dyna_approx_mA} and \ref{Eq:dyna_approx_mB}), we get
\begin{align}	
	\partial_t r =& \bigg(b\bigg(\frac{J}{\sigma}\bigg)\bigg)r - \frac{br^3}{8}\bigg(\frac{J}{\sigma}\bigg)^3\bigg(1 + 2\gamma^2+\bigg(\frac{1}{3}-\gamma^2\bigg)\cos(4\theta)
	\notag\\&+\gamma \bigg(1-\frac{\gamma^2}{3}\bigg)\sin(4\theta)\bigg)
	\label{Eq:dyna_approx_r}
\end{align}
\begin{align}	
	\partial_t \theta =& -b\bigg(\frac{K}{\sigma}\bigg) + \frac{br^2}{8}\bigg(\frac{J}{\sigma}\bigg)^3 \bigg(\gamma(2+\gamma^2) + \gamma\bigg(\frac{\gamma^2}{3}-1\bigg)\cos(4\theta)\bigg)\notag\\&+ \bigg(\frac{1}{3}-\gamma^2\bigg)\sin(4\theta)
	\label{Eq:dyna_approx_theta}
\end{align}
where $\gamma = \frac{K}{J}$.
\subsubsection{Boundary between the disordered and the oscillatory phases for small amplitudes}
For a given $K>0$, let the coefficient of $r$ be $\mu = b\big(\frac{J}{\sigma}\big)-1$. We have the following:\\
For $\mu < 0$ (i.e) $\sigma>\sqrt{\frac{2}{\pi}}J$, $r = 0$ is the fixed point that the system spiral towards (a disordered phase).\\
For $\mu > 0$ (i.e) $\sigma<\sqrt{\frac{2}{\pi}}J$, $r = 0$ becomes an unstable fixed point and the set of points with $r = \sqrt{\mu} = \sqrt{b\Big(\frac{J}{\sigma}\Big)-1}$ constitutes a stable limit cycle.\\
Thus, the line $\sigma=\sqrt{\frac{2}{\pi}}J$ marks the boundary between the disordered and the oscillatory phases. Near the transition, the dynamic equations in $\theta$ can be written as 
$$\partial_t \theta \approx -b \bigg(\frac{K}{\sigma}\bigg) + \mathcal{O}\Bigg(b\bigg(\frac{J}{\sigma}\bigg)-1\Bigg)$$
Thus, $$\theta \approx \theta_0 -b \bigg(\frac{K}{\sigma}\bigg)t + \mathcal{O}\Bigg(b\bigg(\frac{J}{\sigma}\bigg)-1\Bigg)$$
Substituting $\theta$ from the above result in the dynamical equation in $r$ with $\theta_0 = 0$, we get
\begin{align}
	\partial_t r = \Bigg(b\bigg(\frac{J}{\sigma}\bigg)-1\Bigg)r 
	&-\frac{br^3}{8}\bigg(\frac{J}{\sigma}\bigg)^3\bigg((1 + 2\gamma^2)\notag\\&+\bigg(\frac{1}{3}-\gamma^2\bigg)\cos\bigg(4b\bigg(\frac{K}{\sigma}\bigg)t\bigg)\notag\notag\\&+\gamma \bigg(1-\frac{\gamma^2}{3}\bigg)\sin\bigg(4b\bigg(\frac{K}{\sigma}\bigg)t\bigg)\bigg) \notag\\&+ \mathcal{O}\Bigg(\bigg(b\bigg(\frac{J}{\sigma}\bigg)-1\bigg) ^{5/2}\Bigg)
\end{align}

\subsubsection{Boundary between the oscillatory and the ordered steady state for small amplitudes}
When $b\big(\frac{K}{\sigma}\big)$ and $b\big(\frac{J}{\sigma}\big)-1$ are comparable, the transition between an oscillatory and an ordered steady state. Let $b\big(\frac{K}{\sigma}\big)\sim \epsilon$, $b\big(\frac{J}{\sigma}\big)-1\sim \epsilon$. Keeping terms to the lowest order of $\epsilon$ we have 
$$\partial_t r = \bigg(b\bigg(\frac{J}{\sigma}\bigg)-1\bigg)r - \frac{r^3}{8b^2}\bigg(1+\frac{\cos(4\theta)}{3}\bigg)$$
where each term is of $\mathcal{O}(\epsilon^{3/2})$.
$$\partial_t \theta = -b\bigg(\frac{K}{\sigma}\bigg)+\frac{r^2}{24b^2}\sin(4\theta)$$
where each term is of $\mathcal{O}(\epsilon)$.\\
For the ordered steady state near the transition, $\partial_t r = \partial_t \theta = 0$. This gives $$r = \sqrt{\frac{8b^2\Big(b\big(\frac{J}{\sigma}\big)-1\Big)}{1 + \frac{\cos(4\theta)}{3}}}$$ and thus
$$\partial_t \theta = -b\bigg(\frac{K}{\sigma}\bigg)+\frac{\sin(4\theta)}{3+\cos(4\theta)}\Bigg(b\bigg(\frac{J}{\sigma}\bigg)-1\Bigg)$$
Now since $\frac{\sin(4\theta)}{3+\cos(4\theta)} \in \Big[-\frac{1}{\sqrt{2}},\frac{1}{\sqrt{2}}\Big]$, we have $\partial_t\theta \in \Bigg[-b\big(\frac{K}{\sigma}\big)-\frac{b\big(\frac{J}{\sigma}\big)-1}{\sqrt{2}},-b\big(\frac{K}{\sigma}\big)+\frac{b\big(\frac{J}{\sigma}\big)-1}{\sqrt{2}}\Bigg]$. The solution $\partial_t \theta = 0$ is possible only if  $b\big(\frac{K}{\sigma}\big) < \frac{b\big(\frac{J}{\sigma}\big)-1}{2\sqrt{2}}$ and thus the line $K = \frac{J-\sqrt{\frac{\pi}{2}}\sigma}{2\sqrt{2}}$ marks the boundary between the oscillatory and the ordered steady-state phases.\\
Fixed points at the point of transition for $\theta$ that maximize $\frac{\sin(4\theta)}{3+\cos(4\theta)}$ are given by:
$$\theta_n = \frac{1}{4}\Big(-\text{tan}^{-1}\big(2\sqrt{2}+\pi (2n+1)\big)\Big)$$ with $n = 0,1,2,...$. The ordered steady-state magnetisation values are given by:
$$(m_A, m_B) = 2b\left(\sqrt{b\bigg(\frac{J}{\sigma}\bigg)-1}\right)\big(\cos\theta_n, \sin\theta_n\big)$$

Since $\partial_t\theta\rightarrow 0$ as $\theta \rightarrow \theta_n$, oscillation period diverges at the point of transition.\\
Presence of these solutions prove the setting of long range oscillatory order due to broken time translation symmetry.

\subsubsection{Determining the full $K-\sigma$ boundary between oscillatory and ordered phase}
\begin{figure*}[t]
	\centering
	\foreach \a in {1.13, 1.14} {
		\begin{subfigure}{0.48\linewidth}
			\includegraphics[width=\linewidth]{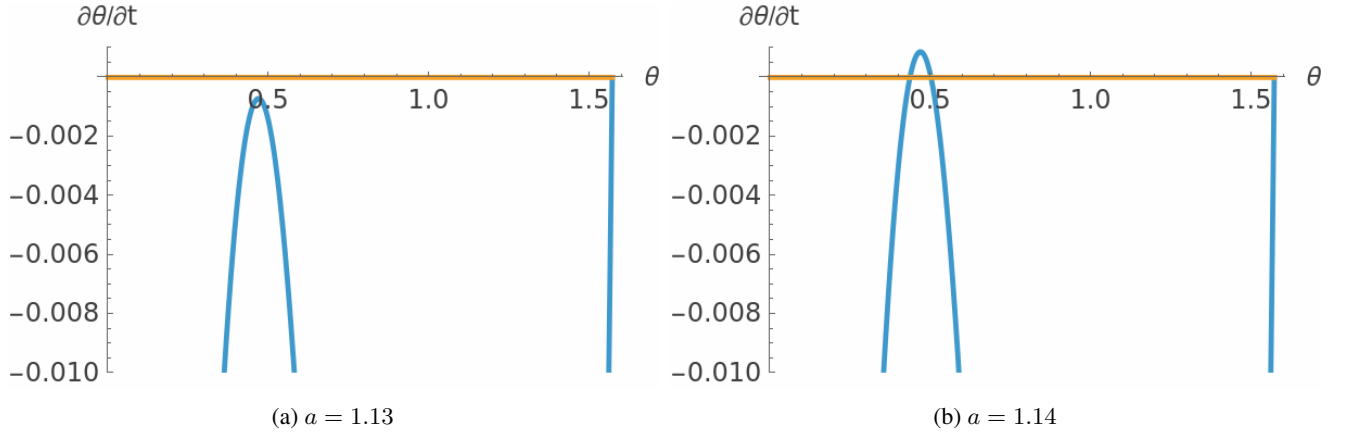}
			\subcaption{\centering $a = \a$}
		\end{subfigure}
	}
	\caption{Plot of $\partial_t \theta$ (from Eq. \ref{partial_theta_2}) as  a function of $\theta$ for $J = 1; K = 0.1$. We see that there is a critical value of $a_c$ such that for $a > a_c$, $\partial_t \theta$ has non-trivial zeros.}
	\label{fig:partial_t_theta_J1_K0.1_a}
\end{figure*}

\begin{figure*}[t]
	\centering
	\includegraphics[width=0.75\textwidth]{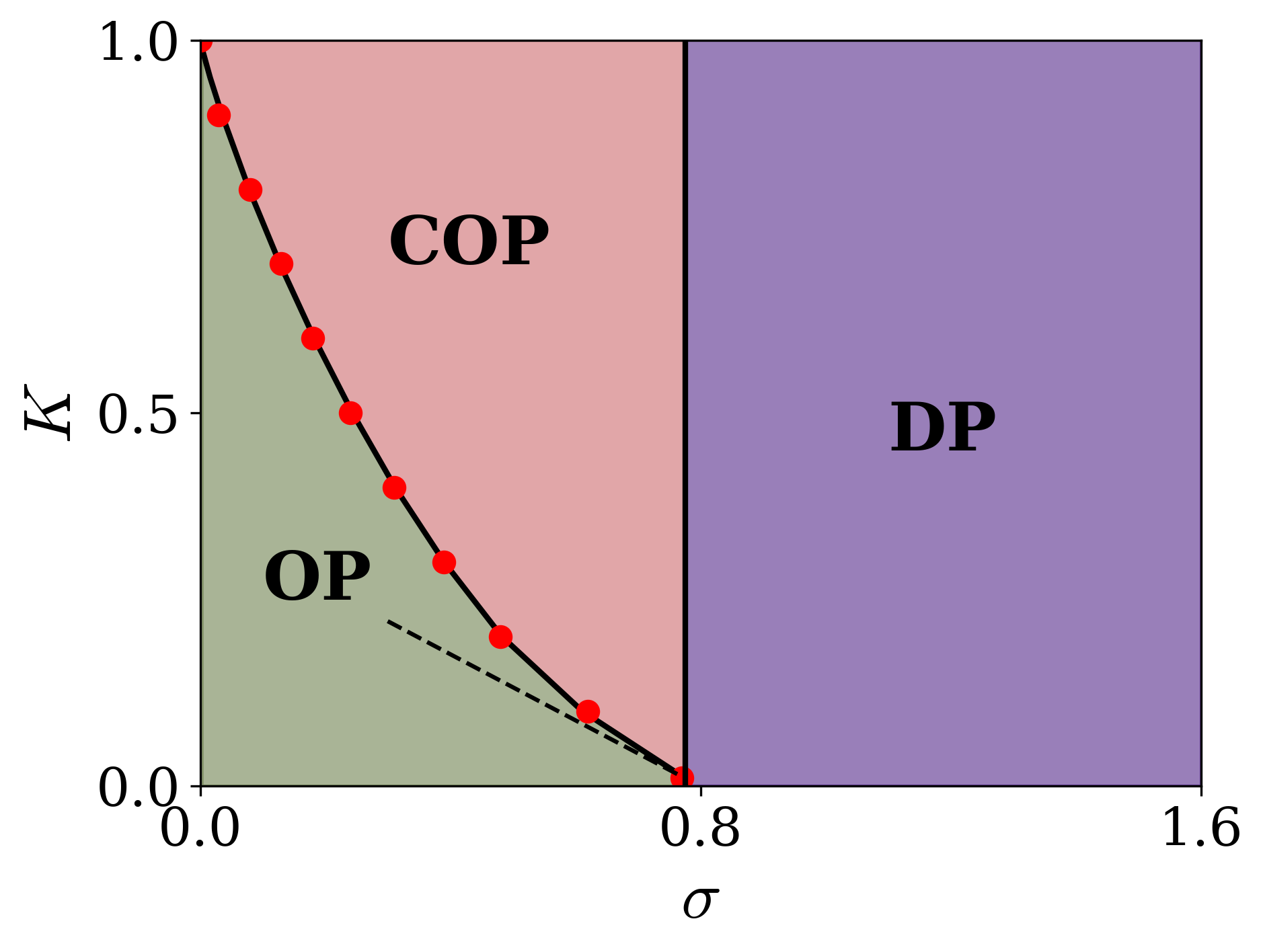}
	\caption{The phase diagram in the $K-\sigma$ plane ($J = 1$) by solving the dynamical equations numerically with the Ordered Phase (OP), the Chaotic Ordered Phase (COP) and Disordered Phase (DP). The solid lines mark the lines of transition between the two phases. The dotted lines denote the linearized calculation of the line of transition between an ordered steady-state and the oscillatory phase: $K = \frac{J-\sqrt{\pi/2}\ \sigma}{2\sqrt{2}}$. The red dots represent the boundary between the phase with non-trivial fixed point and the phase with only $(0,0)$ fixed point of Eqs. \ref{Eq:MF_fp_m_A} and \ref{Eq:MF_fp_m_B}.}
	\label{fig:PD_Ksigma}
\end{figure*}

Using the flow equations in polar co-ordinates (Eqs. \ref{partial_t_r} and \ref{partial_t_theta}), with the condition that the system reaches a steady state (i.e) $\partial_t r = 0$ and $\partial_t \theta = 0$ with $r\neq0$, we have the fixed point equations in the polar co-ordinates:
$$r = X\cos\theta+Y\sin\theta$$
$$\theta = \text{tan}^{-1}\Big(\frac{Y}{X}\Big)$$
We can hence take $	X = r\cos\theta$ and $Y = r\sin\theta$ and using Eqs. \ref{partial_t_theta}, we get
\begin{equation}
	\partial_t \theta = \bigg(\frac{Y}{X}\cos\theta - \sin\theta \bigg)\cos\theta
	\label{partial_theta_2}
\end{equation}
Since the condition $\partial_t r = 0$ is ingrained in the above equation. Thus, the condition that an ordered steady state is reached is given setting $\partial_t \theta = 0$ in Eq. \ref{partial_theta_2}:
\begin{equation}
	\bigg(\frac{Y}{X}\cos\theta - \sin\theta \bigg)\cos\theta = 0
	\label{partial_theta_2_eq_0}
\end{equation}
As we can see Eq. \ref{partial_theta_2_eq_0} always has a solution since the zeros of $\cos\theta$ are its solutions. We hence fix $J$ and $K$ and look for the values of $a (= \frac{r}{\sqrt{2}\sigma})$ for which there exists a non-trivial solution of $\theta$. As shown in Fig. \ref{fig:partial_t_theta_J1_K0.1_a}, there is a critical value of $a_c$ such that for $a > a_c$, $\partial_t \theta$ has non-trivial zeros. The existence of the non-trivial solution was checked numerically using mathematica and the value of $\theta$ at which this happens, $\theta_c$ was also obtained. The values of $a_c$, $\theta_c$, $\sigma_c$ for few values of $K$ for $J = 1$ are tabulated in Table \ref{tab:JK_ac}. Using 
$r\cos\theta = \text{\text{erf}}\big(a(J\cos\theta + K\sin\theta)\big)$ and $r = \sqrt{2}\sigma a$ and the known value of $a_c$ and $\theta_c$, we find the value of $\sigma_c$ for a given $J$ and $K$ using: 
\begin{equation}
	\sigma_c = \frac{\text{\text{erf}}\big(a_c(J\cos\theta_c + K\sin\theta_c)\big)}{ \sqrt{2}a_c\cos\theta_c}
\end{equation}
The above line marks the line of transition between an ordered steady-state phase to an oscillatory phase. Along with the line of transition from an oscillatory phase to a disordered phase: $\sigma_c = \sqrt{\frac{2}{\pi}}J$, one can plot the phase diagram $K-\sigma$ (for $J = 1$), as shown in Fig. \ref{fig:PD_Ksigma}.\\
Interestingly, the phase boundary found by solving the Eq. \ref{partial_theta_2_eq_0} matches with the boundary between the phase with non-trivial fixed point and the phase with only $(0,0)$ fixed point, obtained from Eq. \ref{Eq:MF_fp_m_A} and \ref{Eq:MF_fp_m_B}.

\begin{table}[h]
	\centering
	\begin{tabular}{|c|c|c|c|}
		\hline
		$K$ & $a_c$ & $\theta_c$ & $\sigma_c$ \\
		\hline
		0.01 & 0.35 & 0.442594 & 0.77548 \\
		\hline
		0.1 & 1.14 & 0.501094 & 0.611144 \\
		\hline
		0.2 & 1.69 & 0.559784 & 0.48258 \\
		\hline
		0.3 & 2.24 & 0.631071 & 0.390265 \\
		\hline
		0.4 & 2.9 & 0.673578 & 0.311957 \\
		\hline
		0.5 & 3.82 & 0.707932 & 0.243655 \\
		\hline
		0.6 & 5.22 & 0.735334 & 0.182659 \\
		\hline
		0.7 & 7.61 & 0.755603 & 0.12766 \\
		\hline
		0.8 & 12.56 & 0.768924 & 0.078338 \\
		\hline
		0.9 & 28.37 & 0.779203 & 0.0350321 \\
		\hline
		0.95 & 62.22 & 0.782593 & 0.0160271 \\
		\hline
		0.99 & 364.26 & 0.785009 & 0.00274422 \\
		\hline
		
	\end{tabular}
	\caption{The values of $a_c$, $\theta_c$, and $\sigma_c$ for a given $K$ for $J = 1$ calculated numerically.}
	\label{tab:JK_ac}
\end{table}

\pagebreak

\subsubsection{Phase diagram of $\widetilde{J}-\widetilde{K}$ via numerical analysis}
\begin{figure*}[t]
	\centering
	\includegraphics[width=0.75\textwidth]{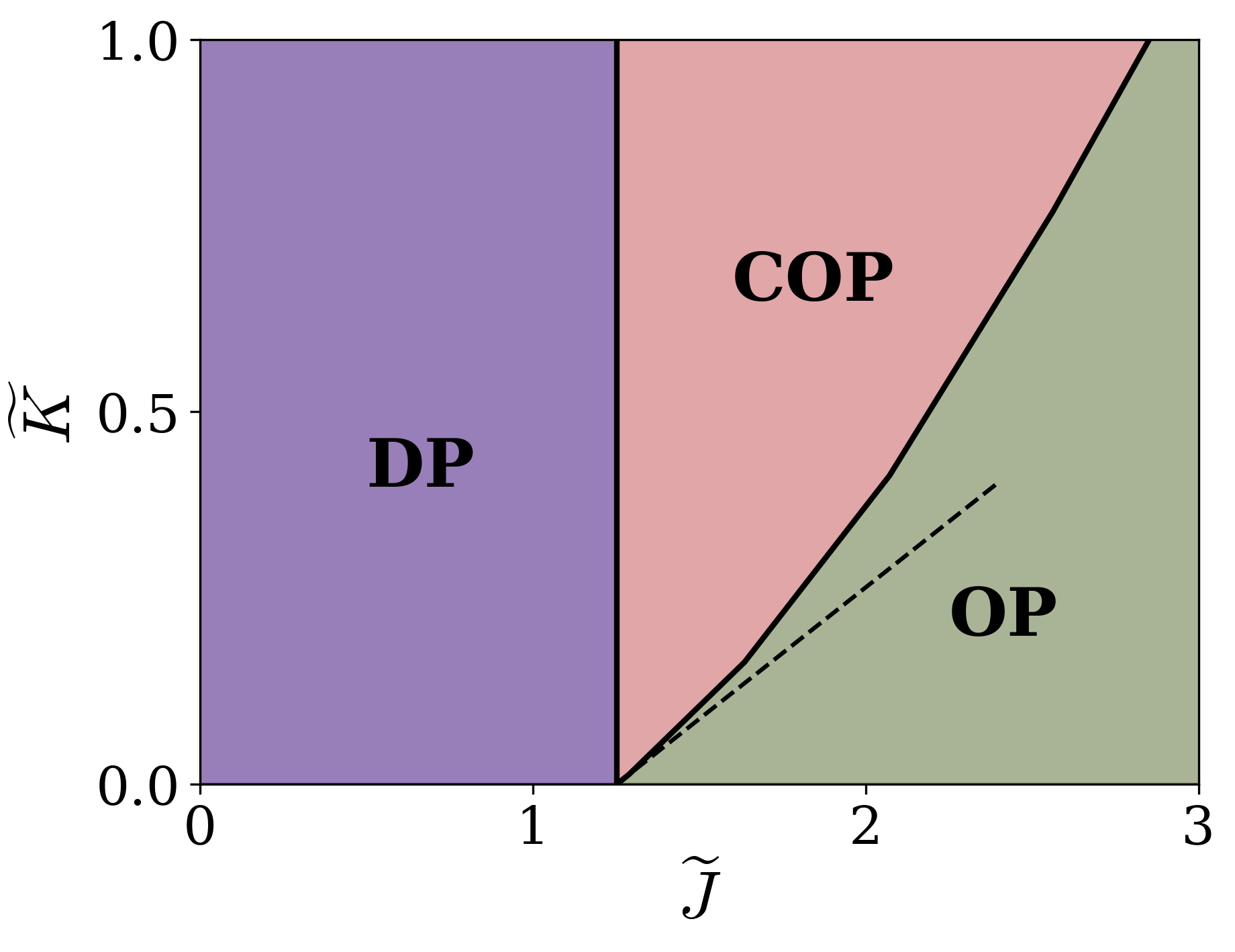}
	\caption{The phase diagram in the $\widetilde{J}-\widetilde{K}$ plane by solving the dynamical equations numerically. There are three phases: the Ordered Phase (OP), the Chaotic Ordered Phase (COP) and Disordered Phase (DP). The solid lines mark the lines of transition between the two phases. The dotted lines denote the linearized calculation of the line of transition between an ordered steady-state and the oscillatory phase: $\widetilde{K} = \frac{\widetilde{J}-\sqrt{\pi/2}}{2\sqrt{2}}$.}
	\label{fig:PD_J_tilde_K_tilde}
\end{figure*}

Defining $\widetilde{J} = \frac{J}{\sigma}$ and $\widetilde{K} = \frac{K}{\sigma}$, we can also obtain $\widetilde{J}-\widetilde{K}$ phase diagram. The equations become: 
\begin{equation}
	\partial_t \theta = \Big(\frac{Y}{X}\cos\theta - \sin\theta \Big)\cos\theta
\end{equation}
where \begin{equation}
	X = \text{\text{erf}}\bigg(\frac{r}{\sqrt{2}}\left(\widetilde{J}\cos\theta + \widetilde{K}\sin\theta\right)\bigg)
	\label{Eq:X2}
\end{equation}
\begin{equation}
	Y = \text{\text{erf}}\bigg(\frac{r}{\sqrt{2}}\left(\widetilde{J}\sin\theta - \widetilde{K}\cos\theta\right)\bigg)
	\label{Eq:Y2}
\end{equation}
The resultant phase diagram is drawn in Fig. \ref{fig:PD_J_tilde_K_tilde}. This qualitatively matches with the corresponding $\widetilde{J}-\widetilde{K}$ phase diagram in \cite{avni2}.

\section{Monte Carlo simulation plots of the order parameters}

Via Monte Carlo simulations we study the following two order parameters: 

\begin{equation}
	R = \left\langle \sqrt{\frac{m_A^2+m_B^2}{2}} \right\rangle_{t,Q}
	\label{Eq:R}
\end{equation}
\begin{equation}
	\mathcal{L} =\left\langle l\right\rangle_{t,Q}
	\label{Eq:L}
\end{equation}
 where the average is performed over time $t$ and $Q$ disorder realizations and $l(t) = m_B(t)\partial_t m_A(t) - m_A(t)\partial_t m_B(t)= m_A(t)m_B(t-1)-m_B(t)m_A(t-1)$ for a given realization. The order parameter $R$ is a marker for ferromagnetic order in the system and its value is high for an ordered phase and low for a disordered phase. The phase space angular momentum $\mathcal{L}$ becomes non-zero in the presence of long-range collective oscillations.\\
The system is prepared in an initial state with a spin configuration that pertains to magentisations $m_A, m_B = +1,+1 \text{ or} +1, -1$. The spin-dynamics is as follows: We label the sites from $1$ to $N$ (where $N$ is the total number of spins). We start by relaxing a spin of one of the species, at random, on the $1^{st}$ site, followed by the $2^{nd}$ site, and so on until the $N^{th}$ site. Two sweeps comprise one Monte Carlo run (one time-step). After a time of transience, the values of the order parameters $R$ and $\mathcal{L}$ obtained after each Monte Carlo run, are considered for calculating the average.
\subsection{Results from simulation}
\begin{figure*}[ht]
	\centering
	\includegraphics[width=0.9\textwidth]{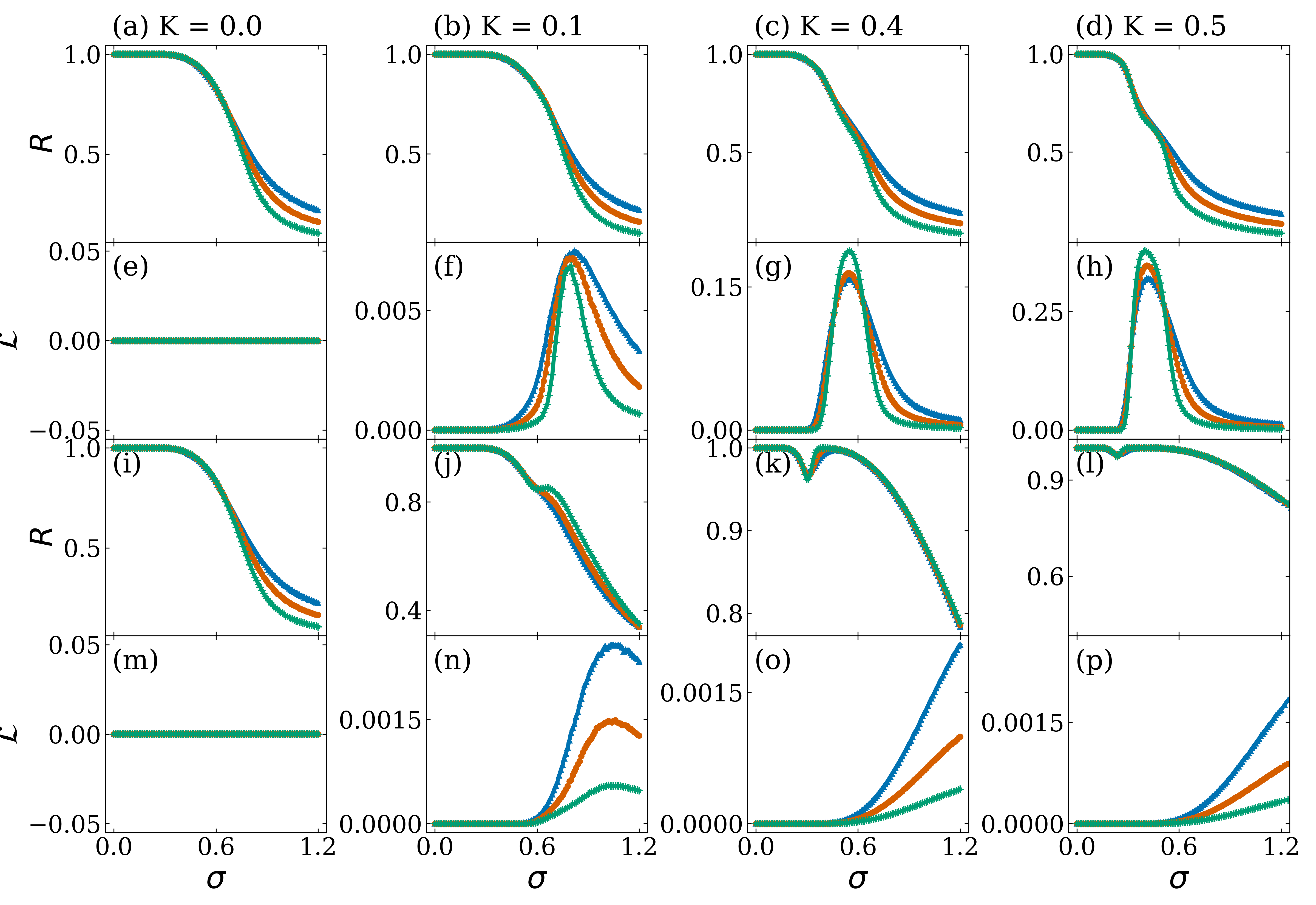}
	\caption{Complete graph with $J = 1$: The plots of order parameter $R$ and $\mathcal{L}$ as a function of $\sigma$ for different system sizes $N = 100, 200$ and $500$ (denoted in blue, orange, and green respectively), averaged over $5\times10^4$, $3\times 10^4$ and $10^4$ realizations of random field distributions respectively. The sub-figures (a)-(h) and (i)-(p) correspond to the initial states $(m_A,m_B)=(+1,+1)$  and $(+1,-1)$ respectively. }
	\label{fig:plot_nr_FCG_ofv_is}
\end{figure*}

\subsubsection{Complete graph (Fig. \ref{fig:plot_nr_FCG_ofv_is})}
The phase diagram obtained in Fig. \ref{fig:PD_Ksigma}, is verified from simulations on a complete graph. For $K = 0$, as $\sigma$ increases, there is a phase transition from a ferromagnetic phase to a paramagnetic phase in all lattices. For a given $K>0$, as we increase $\sigma$ we see that the system goes from Ordered Phase (OP)  (marked by high value of $R$ and low values of $\mathcal{L}$), to a Chaotic Ordered Phase (COP) (marked by high values of $\mathcal{L}$ with non-zero $R$) and finally to a Disordered Phase (DP) (marked by low values of $R$ and $\mathcal{L}$). For $K> 1$ the ordered phase vanishes whereas for $K<1$, DP to COP transition is independent of $\sigma$. The plots of $R$ and $L$ also show dependence on the initial state. For the case of the initial state $(+1,-1)$, $R$ is a non-monotonic function of $\sigma$ for $K = 0.2, 0.3, 0.4, 0.5$. This surprisingly seem to suppress the oscillations. 

\begin{figure*}[ht]
	\centering
	\includegraphics[width=0.9\textwidth]{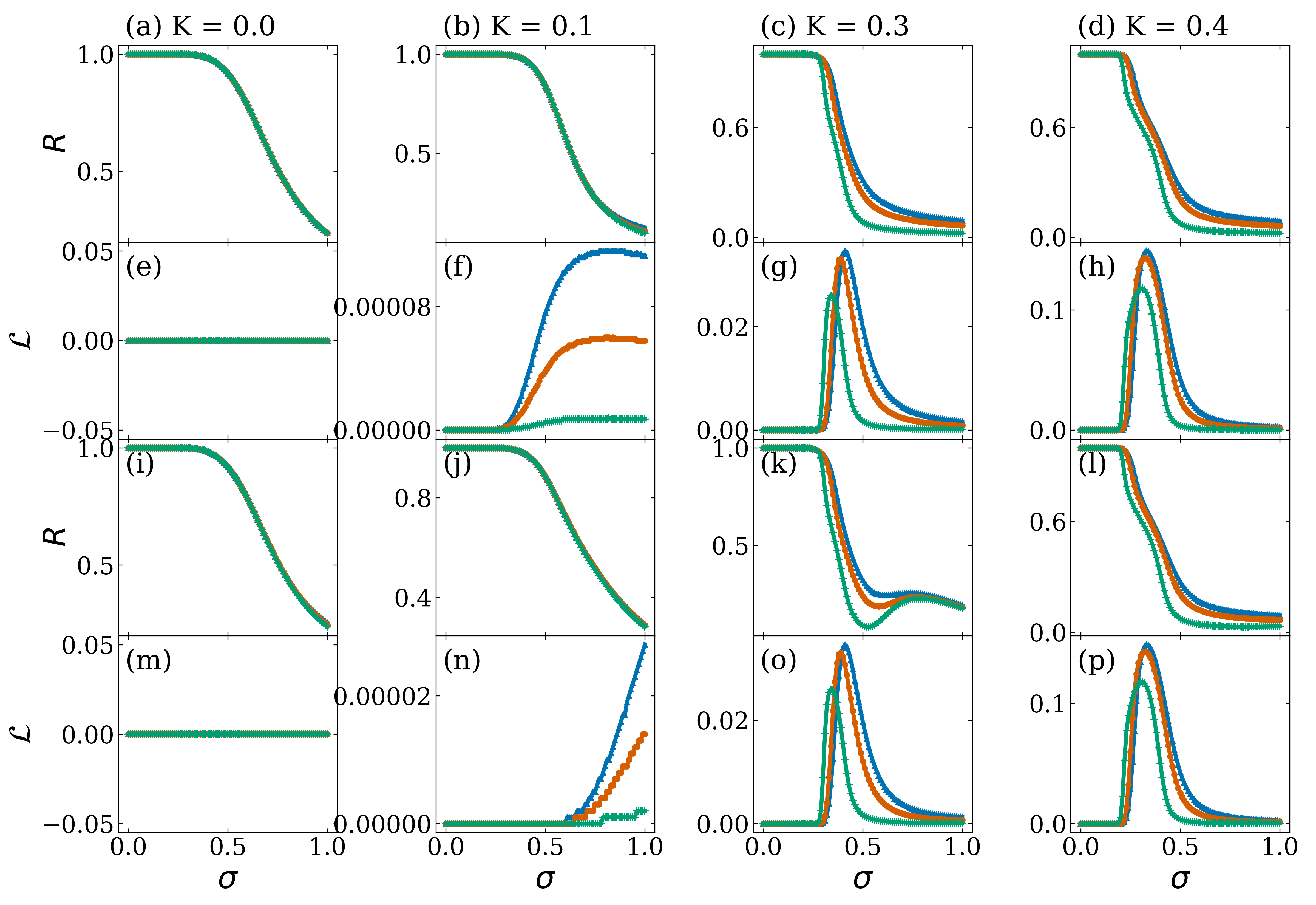}
	\caption{Cubic lattice with $J = 1$: The plots of order parameter $R$ and $\mathcal{L}$ as a function of $\sigma$ for different system sizes $L = 8, 10$ and $20$ (denoted in blue, orange, and green respectively), averaged over $4\times 10^4$, $2\times 10^4$ and $10^4$ realizations of random field distributions respectively. The sub-figures (a)-(h) and (i)-(p) correspond to the initial states $(m_A,m_B)=(+1,+1)$ and $(+1,-1)$ respectively. }
	\label{fig:plot_nr_CL_ofv_is}
\end{figure*}

\subsubsection{Cubic lattice (Fig. \ref{fig:plot_nr_CL_ofv_is})}
Similar to the case of a complete graph, as $\sigma$ increases, for $K = 0$, a phase transition from an ordered phase to a disordered phase occurs and unlike the case of a complete graph, the system is not sensitive to the initial conditions in this case. 

\begin{figure*}[ht]
	\centering
	\includegraphics[width=0.9\textwidth]{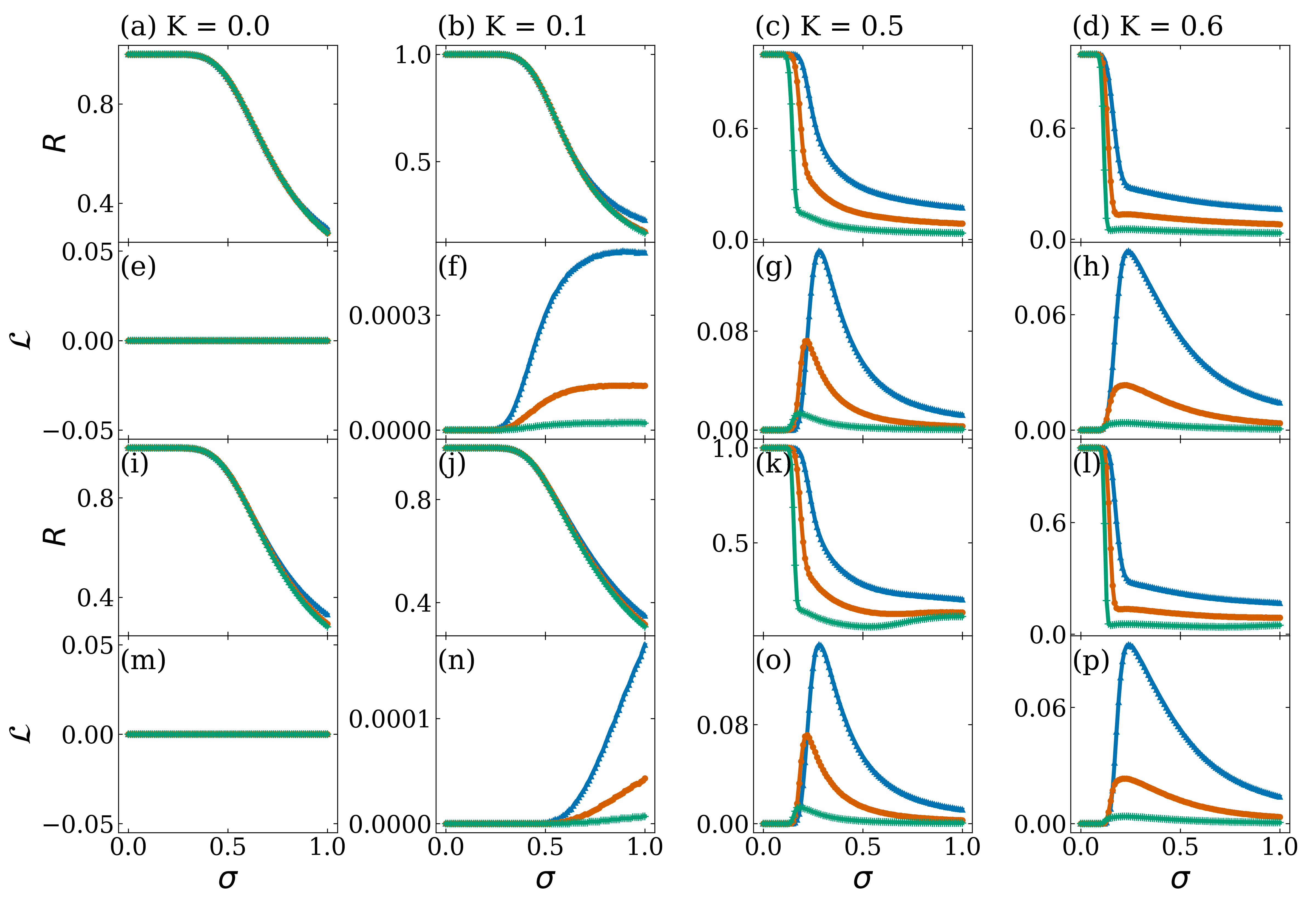}
	\caption{Square lattice with $J = 1$: The plots of order parameter $R$ and $\mathcal{L}$ as a function of $\sigma$ for different system sizes $L = 10, 20$ and $50$ (denoted in blue, orange, and green respectively), averaged over $10^5$, $2\times 10^4$ and $10^4$ realizations of random field distributions respectively. The sub-figures (a)-(h) and (i)-(p) correspond to the initial states $(m_A,m_B)=(+1,+1)$ and $(+1,-1)$ respectively. }
	\label{fig:plot_nr_SL_ofv_is}
\end{figure*}

\subsubsection{Square lattice (Fig. \ref{fig:plot_nr_SL_ofv_is})}
For $K = 0$, the system has a transition around $\sigma \approx 0.64$. This is consistent with the reported non-equilibrium steady-state transition for zero Temperature RFIM on a square lattice \cite{frontera}. We find that $\mathcal{L}$ is highly suppressed on a square lattice.


\section{Time oscillations in order parameter and energy}
\subsection{Time plots for a single realization}
\begin{figure*}[ht]
	\centering
	\includegraphics[width=0.95\textwidth]{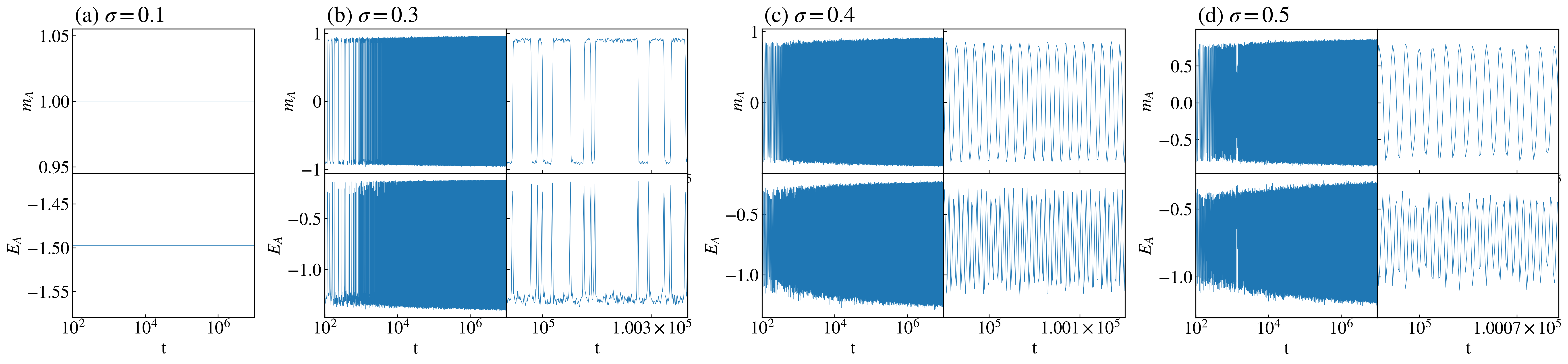}
	\caption{$m_A$ and $E_A$ as a function of time $t$ of a complete graph for $N = 1000$ with $J = 1$ for $K = 0.5$. }
	\label{fig:plot_t_crystal_FCG}
\end{figure*}

\begin{figure*}[ht]
	\centering
	\includegraphics[width=0.8\textwidth]{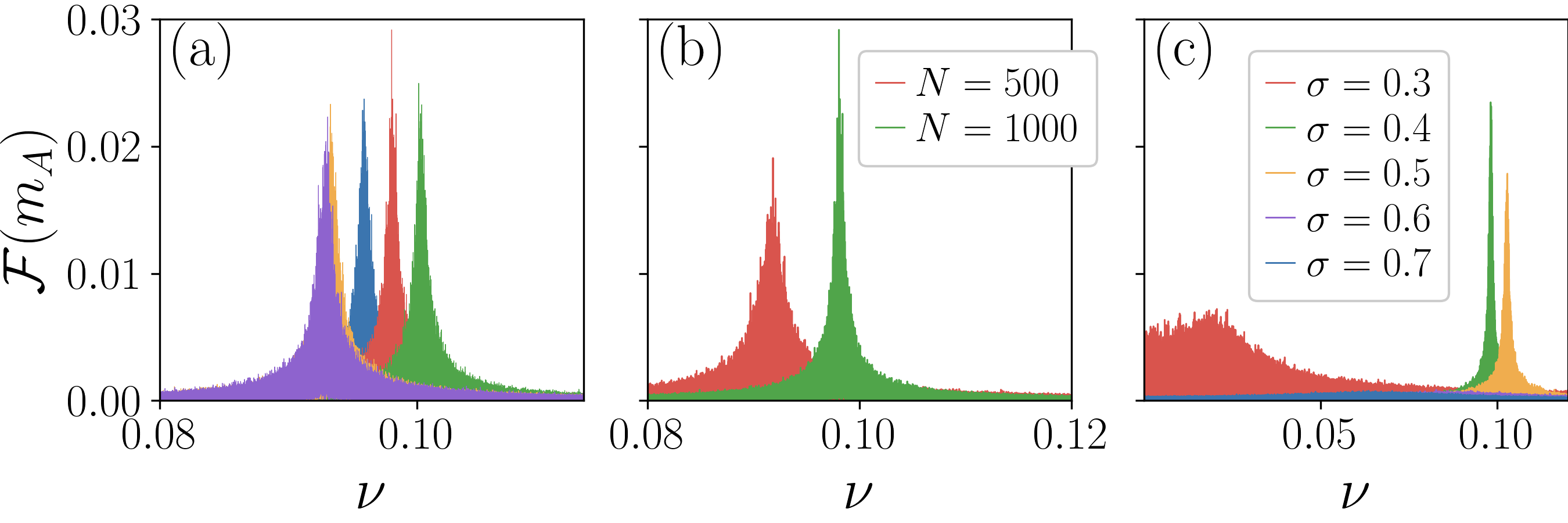}
	\caption{Plots of Fourier transform ($\mathcal{F}(m_A)$) of $m_A(t)$ as a function of frequency $\nu$ for a complete graph: a) shows the plots of $N = 1000$ with $K = 0.5;\sigma = 0.4$ obtained from five different realizations of random fields. b) shows the plots for two system sizes $N = 500, 1000$ with $K = 0.5;\sigma = 0.4$. c) shows plots of different values of $\sigma$, for $N = 1000$ with $K = 0.5$.}
	\label{fig:plot_corr_freq_FCG}
\end{figure*}

\begin{figure*}[ht]
	\centering
	\includegraphics[width=0.95\textwidth]{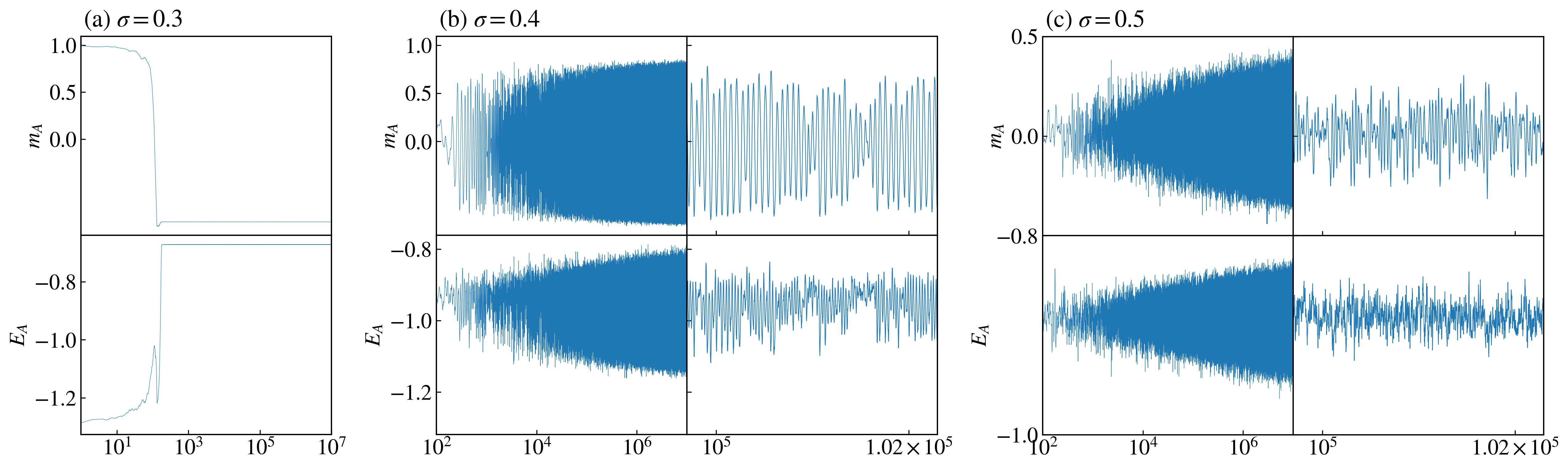}
	\caption{$m_A$ and $E_A$ as a function of time $t$ of a cubic lattice of $L = 20$ with $J = 1$ for $K = 0.3$}
	\label{fig:plot_t_crystal_CL}
\end{figure*}

\begin{figure*}[ht]
	\centering
	\includegraphics[width=0.8\textwidth]{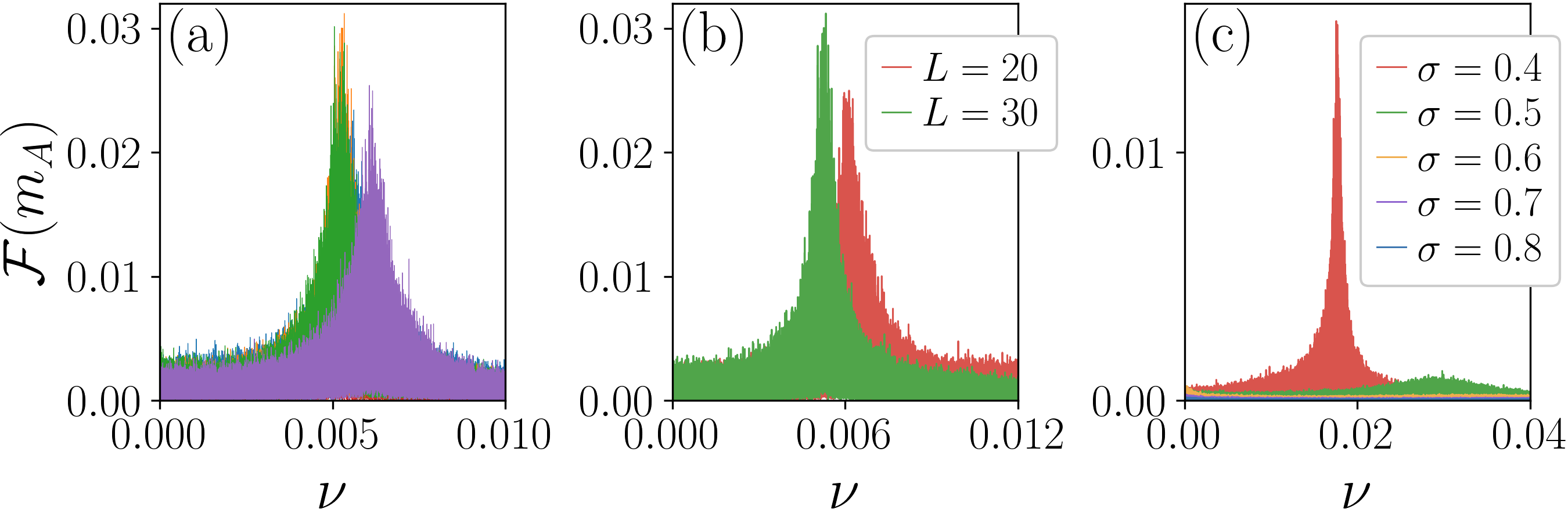}
	\caption{Plots of Fourier transform ($\mathcal{F}(m_A)$) of $m_A(t)$ as a function of frequency $\nu$ for a cubic lattice: a) shows the plots of $L = 30$ with $K = 0.3;\sigma = 0.3$ obtained from five different realizations of random fields. b) shows the plots for two system sizes $L = 20,30$ with $K = 0.3;\sigma = 0.3$. c) shows plots of different values of $\sigma$, for $L = 20$ with $K = 0.3$.}
	\label{fig:plot_corr_freq_CL}
\end{figure*}

We plot here the long-time oscillations in magnetisation  ($m_A$) and energy ($E_A$) as functions of time and the plots of the Fourier transform ($\mathcal{F}(m_A)$) of $m_A(t)$ for a complete graph and cubic lattice (Figs. \ref{fig:plot_t_crystal_FCG} - \ref{fig:plot_corr_freq_CL}). Similar behaviour is also seen for $m_B$ and $E_B$. The $\mathcal{F}(m_A)$ plots show the existence of a certain characteristic frequency ($\nu_0$) that is predominant in the $m_A$ oscillations. However, as seen in Fig. \ref{fig:plot_corr_freq_FCG}(a) and \ref{fig:plot_corr_freq_CL}(a), the value of $\nu_0$ varies with different realizations of random fields. We chose the realization that gave the most prominent peak for a given system size and plotted Fig. \ref{fig:plot_corr_freq_FCG}(b) and \ref{fig:plot_corr_freq_CL}(b) for two different system sizes. The peak is sharper and higher for the larger system size. Moreover, when, for a fixed $K$, we plot $\mathcal{F}(m_A)$ for different values of $\sigma$, the peak is sharper for a range of $\sigma$. The frequency plots also get sharper with increasing system size as shown in Fig.  \ref{fig:plot_corr_freq_FCG}(b) and \ref{fig:plot_corr_freq_CL}(b).

\subsection{Autocorrelation function in time}
\begin{figure*}[t]
	\centering
	
	\begin{subfigure}{0.8\linewidth}
		\includegraphics[width=\linewidth]{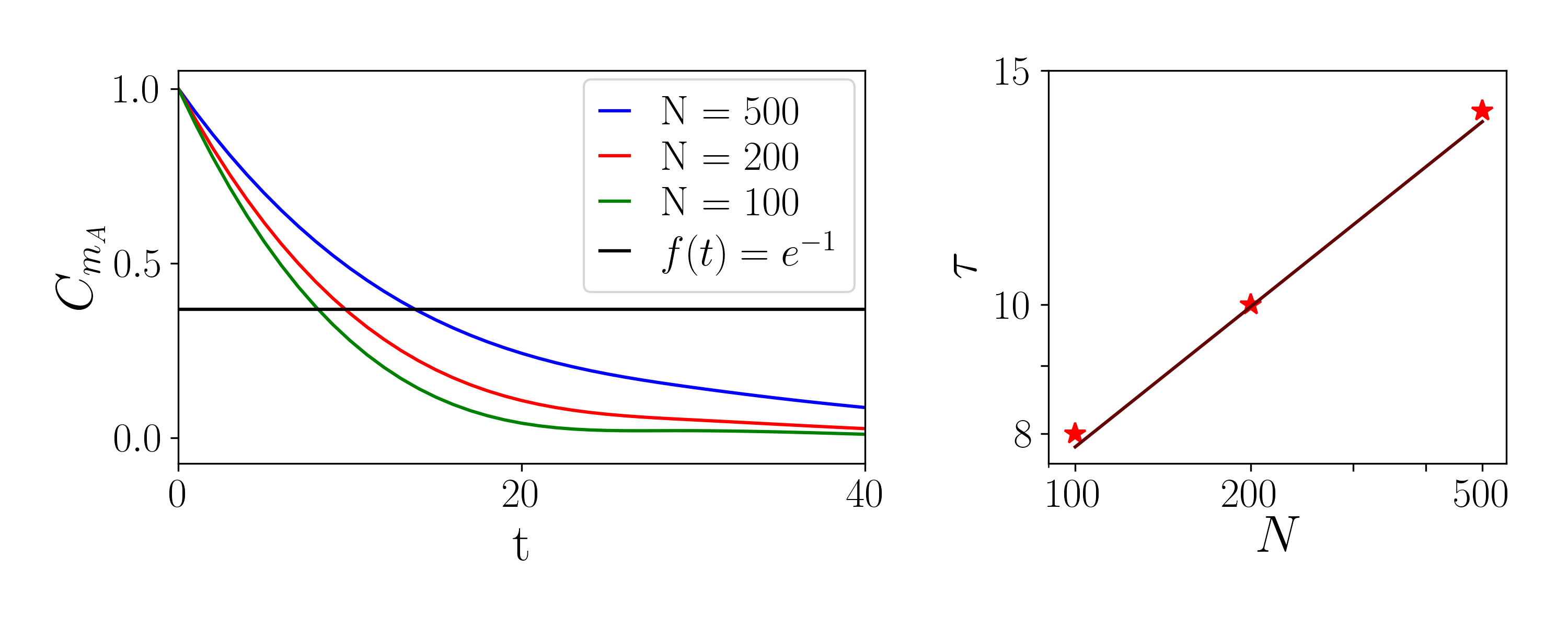}
		\subcaption{\centering Fully connected graph: $J = 1; K = 0.4; \sigma = 0.5$}
	\end{subfigure}

	\begin{subfigure}{0.8\linewidth}
		\includegraphics[width=\linewidth]{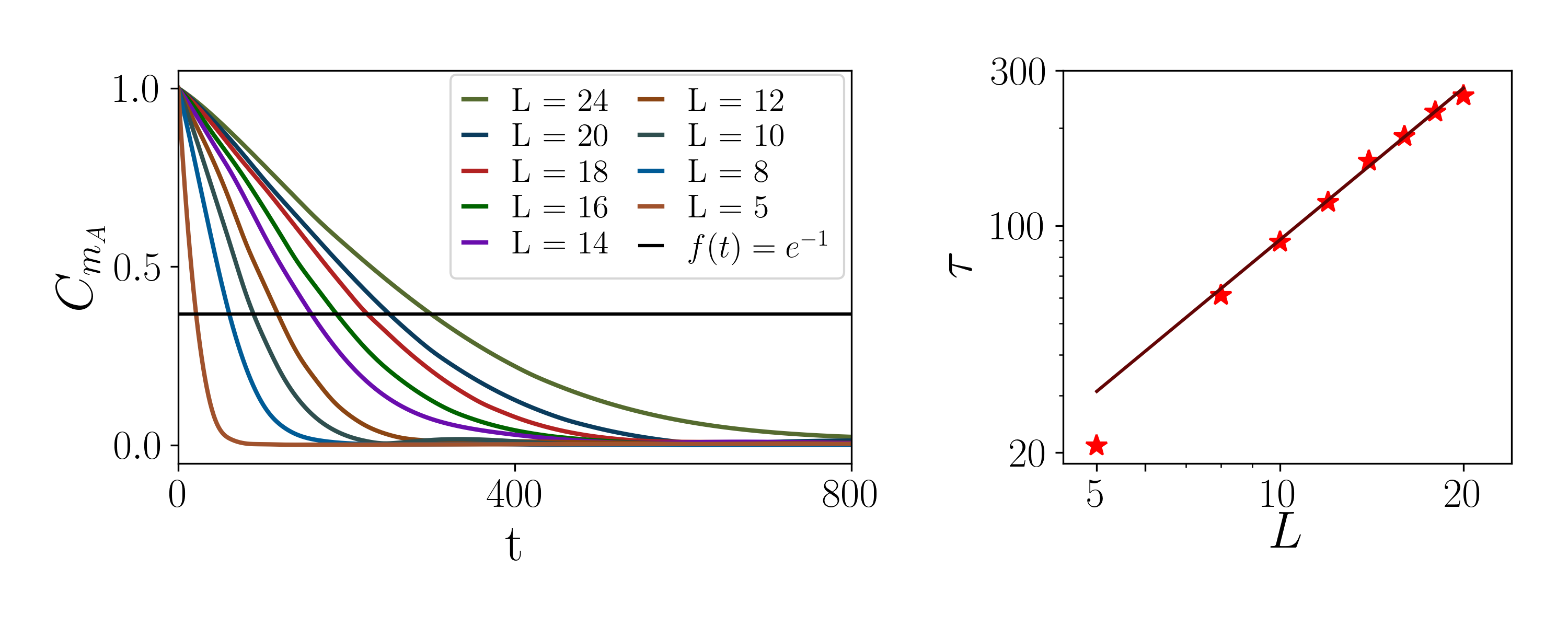}
		\subcaption{\centering Cubic Lattice: $J = 1; K = 0.4; \sigma = 0.4$}
	\end{subfigure}
	\begin{subfigure}{0.8\linewidth}
	\includegraphics[width=\linewidth]{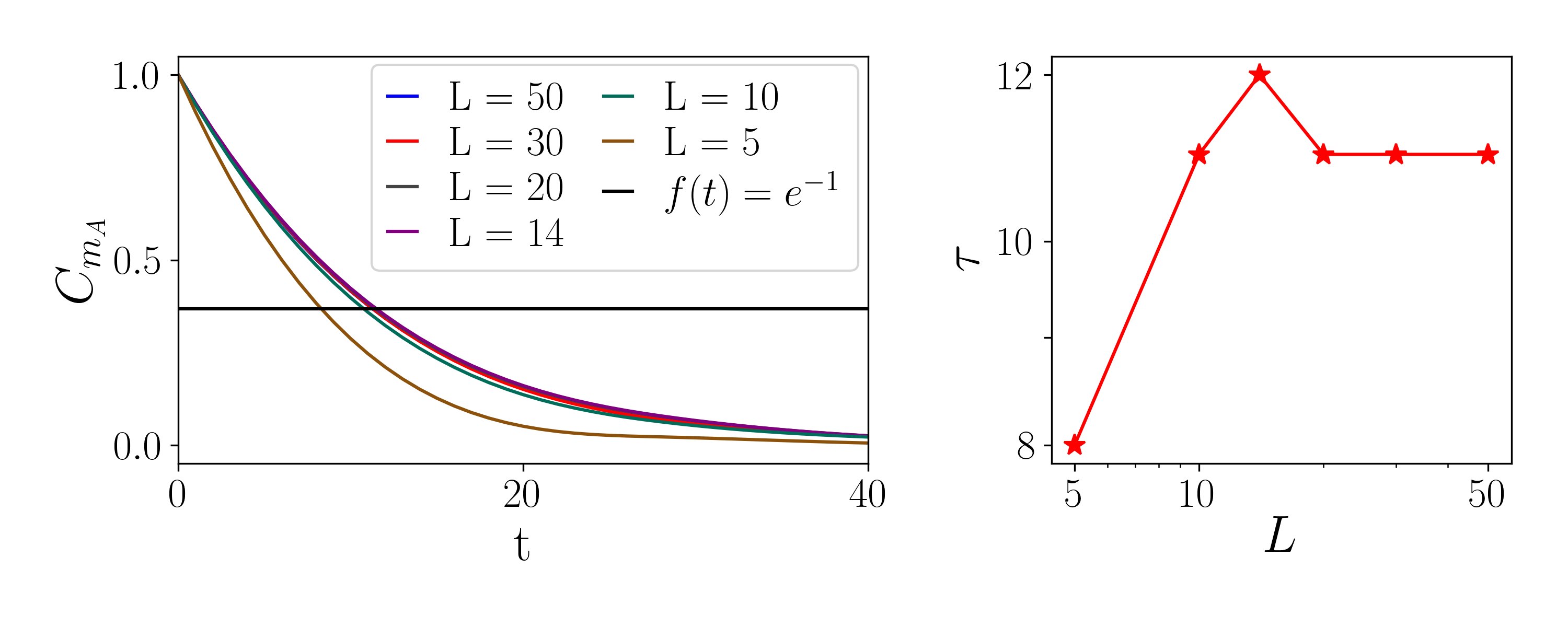}
	\subcaption{\centering Square lattice: $J = 1; K = 0.5; \sigma = 0.4$}
\end{subfigure}
	\caption{The envelope of the oscillations of the time autocorrelation functions of $m_A$ and the value of correlation time $\tau$ found for the case of the complete graph, Cubic lattice and Square lattice. Both complete graph and the cubic lattice show the scaling of $\tau$ as a function of system size. The values of $\tau$ obtained for each system size $S$ was fitted with the function $f(S) = aS^{b\tau}$. For the fully connected graph and cubic lattice, a good fit is obtained for the values $a = 1.56\pm 0.11; b = 0.35\pm 0.01$ and $a = 2.55 \pm 0.39; b = 1.55 \pm 0.06$ respectively.}
	\label{fig:plot_envelope_corr_t}
\end{figure*}

The autocorrelation function in time of a quantity $x$ is defined as follows:
$$C_x(t) = \frac{\Big\langle \big(x(t^\prime)- \langle x(t^\prime)\rangle\big) \big(x(t^\prime+t)- \langle x(t^\prime)\rangle\big) \Big\rangle_{t^\prime,Q}}{\Big\langle \big( x(t^\prime) - \langle x(t^\prime)\rangle \big)^2\Big\rangle_{t^\prime,Q}}$$

where $x(t^\prime)$ is the value of $x$ at time $t^\prime$ and $ \langle ...\rangle$ is the average over different realizations of random fields, for many different runs and for different values of $t^\prime$ in the same run. \\
The correlation time ($\tau$) can be found in two ways:
\begin{itemize}
	\item By identifying the time $t$ at which the upper envelope of the oscillations intersects with $e^{-1}$
	\item By calculating $\tau = \frac{1 }{2}+ \sum_{t=0}^{\infty} |C_{x}(t)|$ 
\end{itemize}
We find $\tau$ for different system sizes using both the methods. There is a good match in the value of $\tau$ found from the two methods. However, since simulations are for finite time, we use the first definition. The results are shown in Fig. \ref{fig:plot_envelope_corr_t}.

\end{document}